\documentclass[11pt]{article}

\usepackage{graphicx, amssymb, amsmath, xcolor, cite}
\usepackage{paralist} 
\usepackage{multirow, booktabs, colortbl, tabularx, array} 
\usepackage{rotating} 

\DeclareFontFamily{OMX}{MnSymbolE}{}
\DeclareFontShape{OMX}{MnSymbolE}{m}{n}{
    <-6>  MnSymbolE5
   <6-7>  MnSymbolE6
   <7-8>  MnSymbolE7
   <8-9>  MnSymbolE8
   <9-10> MnSymbolE9
  <10-12> MnSymbolE10
  <12->   MnSymbolE12}{}
\DeclareFontShape{OMX}{MnSymbolE}{b}{n}{
    <-6>  MnSymbolE-Bold5
   <6-7>  MnSymbolE-Bold6
   <7-8>  MnSymbolE-Bold7
   <8-9>  MnSymbolE-Bold8
   <9-10> MnSymbolE-Bold9
  <10-12> MnSymbolE-Bold10
  <12->   MnSymbolE-Bold12}{}
\DeclareSymbolFont{largesymbolsX}  {OMX}{MnSymbolE}{m}{n}
\SetSymbolFont{largesymbolsX}{bold}{OMX}{MnSymbolE}{b}{n}
\DeclareMathAccent{\widehat}{\mathord}{largesymbolsX}{'302}
\renewcommand{\div}{\mathrm{div}}
\newcommand{\bs}{\boldsymbol}
\newcommand{\eff}{\mathrm{eff}}
\newcommand{\rf}{\mathrm{opt}}
\newcommand{\diag}{\mathrm{diag}}

\newcommand{\opt}{\mathrm{opt}}
\newcommand{\out}{\mathrm{out}}
\newcommand{\into}{\mathrm{in}}
\newcommand{\tot}{\mathrm{tot}}
\newcommand{\LL}{\bs{\vec L}} 
\newcommand{\II}{{\mathbb I}\hspace{-3pt}{\mathbb I}}
\newcommand{\var}{\mathbb V\hspace{-1pt}\mathrm{ar}}
\newcommand{\cov}{\mathbb C\hspace{-1pt}\mathrm{ov}}
\newcommand{\otop}{_\circ{\!\!\!^\top}}
\newcommand{\opos}{_{\hspace{-1pt}\scriptscriptstyle\oplus}}
\newcommand{\oneg}{_{\hspace{-1pt}\scriptscriptstyle\ominus}}
\newcommand{\onul}{_{\hspace{-1pt}\varnothing}}
\newcommand{\QEDA}{\begingroup\small{\hfill $\blacksquare$}\endgroup}

\usepackage{amsthm}

\theoremstyle{plain}
\newtheorem{theorem}{Theorem}

\newtheorem{corollary}{Corollary}
\newtheorem{lemma}{Lemma}

\theoremstyle{definition}
\newtheorem{definition}{Definition}
\newtheorem{assumption}{Assumption}

\theoremstyle{remark}
\newtheorem{notation}{Notation}
\newtheorem{remark}{Remark}

\begin{document}

\title{Heat-Diffusion: Pareto Optimal Dynamic Routing for Time-Varying Wireless Networks}

\date{\vspace{-5ex}}


\author{
Reza~Banirazi, Edmond~Jonckheere, Bhaskar~Krishnamachari
\thanks{This paper is the full and revised version of the conference paper~\cite{RezaInfocomm}.}
\thanks{This work was supported by NSF grants CNS-1017881 and CNS-1049541.}
\thanks{The authors are with Ming Hsieh Department of Electrical Engineering, University of Southern California, Los Angeles, CA 90089.} 
\thanks{E-mail: \{banirazi, jonckhee, bkrishna\}@usc.edu}
}

\renewcommand\footnotemark{}

\maketitle

\begin{abstract}
A dynamic routing policy, referred to as Heat-Diffusion (HD), is developed for multihop uniclass wireless networks subject to random traffic, time-varying topology and inter-channel interference.
The policy uses only current condition of queue occupancies and channel states, with requiring no knowledge of traffic and topology. 
Besides throughput optimality, HD minimizes an average quadratic routing cost defined by endowing each channel with a time-varying cost factor. 
Further, HD minimizes average network delay in the class of routing policies that base decisions only on current condition of traffic congestion and channel states. 
Further, in this class of routing policies, HD provides a Pareto optimal tradeoff between average routing cost and average network delay, meaning that no policy can improve either one without detriment to the other.
Finally, HD fluid limit follows graph combinatorial heat equation, which can open a new way to study wireless networks using heat calculus, a very active area of pure mathematics. 
\end{abstract}



\section{Introduction}

Throughput optimality, which means utilizing the full capacity of a wireless network, is critical to respond to increasing demand for wireless applications. 
The seminal work in~\cite{Tassiulas92} showed that the link queue-differential, channel rate-based {\it Back-Pressure~(BP)} algorithm is throughput optimal under very general conditions on arrival rates and channel state probabilities.
Follow-up works showed that the class of throughput optimal routing policies is indeed large~\cite{Shah06, Dai08, Ross09, Naghshvar12}. 
The challenge is then to develop one that, in addition, is optimal relative to some other important routing objectives.

We propose Heat-Diffusion (HD), a throughput optimal routing policy that operates under the same general conditions and with the same algorithmic structure, complexity and overhead as BP, while also holding the following important qualities:
(i)~HD minimizes the average quadratic routing cost $\overline{R}$ in the sense of Dirichlet.
Endowing each wireless link with a time-varying cost factor, we define average {\it Dirichlet routing cost} as the product of the link cost factors and the square of the average link flow rates.
Such a generic routing cost may reflect different topology-based penalties, e.g., channel quality, routing distance and power usage, even a cost associated with greedy hyperbolic embedding~\cite{Baras12}.
(ii)~HD minimizes average total queue congestion $\overline{Q}$, which is proportional to average network delay by Little's Theorem, within the class of routing algorithms that use only current queue occupancies and current channel states, possibly together with the knowledge of arrival/channel probabilities.
(iii)~In the same class, HD operates on the {\it Pareto boundary} of performance region built on the average network delay $\overline{Q}$ and the average quadratic routing cost $\overline{R}$ and can be made to move along this boundary by changing a control parameter $\beta$ that compromises between the two objectives $\overline{Q}$ and $\overline{R}$ (see Fig.~\ref{figure1}).


{\it Related works---}The study of BP schemes has been a very active research area with wide-ranging applications and many recent theoretical results.
In packet switches, congestion-based scheduling~\cite{Dai08,Dai05, Leonardi} was extended to admit more general functions of queue lengths with particular interest on $\alpha$-weighted schedulers using $\alpha$-exponent of queue lengths~\cite{Shah06}. 
As another extension in packet switches, \cite{Ross09} introduced Projective Cone Schedulers (PCS) to allow scheduling with non diagonal weight assignments. 
The work in~\cite{Naghshvar12} generalized PCS using a tailored ``patch-work'' of localized piecewise quadratic Lyapunov functions. 

In wireless networks, shadow queues enabled BP to handle multicast sessions with reduced number of actual queues that need to be maintained~\cite{Bui11}. 
Replacing queue-length by packet-age, \cite{Shroff11} introduced a delay-based BP policy. 
To improve BP delay performance, \cite{Huang12} proposed place-holders with Last-In-First-Out (LIFO) forwarding.
Adaptive redundancy was used in~\cite{Alresaini12} to reduce light traffic delay in intermittently connected mobile networks. 
Using graph embedding, \cite{Baras12} combined BP with a greedy routing algorithm in hyperbolic coordinates to obtain a throughput-delay tradeoff.

There have been several reductions of BP to practice in the form of distributed wireless protocols of pragmatically implemented and experimentally evaluated~\cite{Moeller10,Martinez11,Srikant09a}. 
Some attempts have also been made to adopt the BP framework for handling finite queue buffers~\cite{Ekici12}. 

Similar to BP, also HD rests on a centralized scheduling with a computational complexity that can be prohibitive in practice. 
Fortunately, much progress has recently been made to ease this difficulty by deriving decentralized schedulers with the performance of arbitrarily close to the centralized version as a function of complexity~\cite{Lin05, Bui09, Jiang11}.  


{\it Contributions---}We derive HD from combinatorial analogue of classical heat equation on smooth manifolds, which leads to the following key contributions:

{\it (Fluid)} Translating ``queue occupancy measured in packets'' to ``heat quantity measured in calories,'' the {\it fluid limit} of interference HD flow mimics a suitably-weighted non-interference heat flow, in agreement with the second Principle of thermodynamics.
In doing so, we introduce a new paradigm that might be called ``wireless network thermodynamics,'' which builds a rigorous connection between wireless networking and well-studied domains of physics and mathematics.

{\it (Cost)} HD reduces the Dirichlet routing cost to its minimum feasible value among all stabilizing routing algorithms. 
To the best of our knowledge, this is the first time a feasible routing algorithm asserts the strict minimization of a cost function subject to network stability, i.e., bounded queue occupancies and network delay. 
This is while the drift-plus-penalty approach of \cite{NeelyBook}, as the best-known alternative, can get only close to the minimum of this routing cost at the expense of infinitely large network delay.

{\it (Delay)} HD minimizes average queue lengths, and so average network delay, within the class of routing algorithms that act based only on current condition of queue occupancies and channel states, including those with the perfect knowledge of arrival/channel probabilities. 
This important class contains stationary randomized algorithms~\cite{NeelyBook}, original BP policy~\cite{Tassiulas92}, and most BP derivations~\cite{Shah06,Dai08,Ross09,Naghshvar12,Baras12,Dai05,Bui11,Shroff11,Huang12,Alresaini12,Moeller10,Martinez11,Srikant09a,Ekici12,Bui09,Jiang11,Srikant09b}.

{\it (Pareto)} In the class of algorithms defined in (Delay), let the performance region built on average delay and the Dirichlet routing cost be convex. 
Then HD operates on the Pareto boundary of this region while the optimal tradeoff can solely be controlled by a routing parameter independently of network topology and traffic. 
This means that no other policy in this class can make a better compromise between these two routing objectives and that any deviation from HD operation leads to the degradation of at least one of them. 

{\it (Complexity)} Last but not least, HD enjoys the same algorithmic structure, complexity and overhead as BP, giving it the same wide-reaching impact. 
This also provides an easy way to leverage all advanced improvements of BP to further enhance HD quality.
At the same time, it simplifies the way to practice via a smooth software transition from BP to HD.


{\it Continuation---}The infant idea of HD algorithm first appeared in \cite{Reza12}, very different indeed from what is called HD in this paper.
The results on minimum network delay are extended to multiclass wireless networks in~\cite{RezaACC}.
By developing the idea of mapping a wireless network onto a nonlinear resistive network, the results on minimum routing cost are extended to multiclass wireless networks in \cite{RezaACM}.  
By extending the principles of classical thermodynamics to routing and resource allocation on wireless networks, the concept of ``wireless network thermodynamics'' is fully established in \cite{RezaThermo}.


{\it Organization---}After preliminaries in the next section,
HD policy is introduced in Sec.~\ref{s:C3_HDdesign} followed by some illustrative examples.
Section~\ref{s:C3_key} presents HD key property -- a foundation to all HD features.
Using {\it Lyapunov theory,} throughput optimality is proven in Sec.~\ref{s:C3_throughput}. 
We show in Sec.~\ref{s:C3_queue_minimum} that HD minimizes average network delay in a class of routing policies.
Physics-oriented model of heat process on {\it directed graphs} is proposed in Sec.~\ref{s:C3_heat}.
Using {\it fluid limit theory,} Sec.~\ref{s:C3_thermodynamics} shows that in limit, HD packet flow resembles combinatorial heat flow on its underlying directed graph.
Using {\it heat calculus,} Sec.~\ref{s:C3_cost_minimization} shows that HD strictly minimizes the Dirichlet routing cost.
HD {\it Pareto optimal} performance is discussed in Sec.~\ref{s:C3_pareto}. 
The paper is concluded in Sec.~\ref{s:C3_conclusion}. 


{\it Notation---}We denote vectors by bold lowercase and matrices by bold capital letters.
By $\bs 0$ we denote the vector of all zeros, by $\bs 1$ the vector of all ones, and by $\bs I$ the identity matrix.
On arrays: 
$\min$ and $\max$ are taken entrywise;
$\preccurlyeq$ and $\succcurlyeq$ express entrywise comparisons; and
$\odot$ denotes the Schur product.
For $\bs v$ as a vector, $\bs v^\top\!$ denotes its transpose, $\diag(\bs v)$ its diagonal matrix expansion, $\Vert \bs v\Vert$ its Euclidean norm, and $\bs v{^+}\! := \max\{\bs 0,\,\bs v\}$.
For $\cal S$ as a set, $\vert{\cal S}\vert$ denotes its cardinality.
We use $\mathbb I$ as the scalar indicator function, and $\II_{\bs v\succ\bs 0}$ as the vector indicator function that its entry $i$ takes the value 1 if $v_i\!>\!0$, and 0 otherwise.
By $\dot x(t)$ we denote the time derivative of $x(t)$.
For a variable $x$ related to a directed edge~$\ell$ from node~$i$ to node~$j$, we use notations $x_\ell$ and $x_{ij}$ interchangeably.
We use $\into(i)$ and $\out(i)$ to denote the sets of nodes neighbor to node $i$ with respectively incoming links to and outgoing links from node~$i$.


\textit{\textbf{Note:}} 
To keep continuity and enhance readability of the manuscript, proofs are all placed in Appendix.

\section{Preliminaries}
\label{s:C3_preliminaries}

Consider a {\it uniclass} wireless network that operates in slotted time with normalized slots $n \!\in\! \{0,1,2,\cdots\}$.
The network is described by a {\it simple, directed} connectivity graph with set of nodes~$\cal V$ and directed edges~$\cal E$.
New packets, all with the same destination at node $d\in\cal V$, {\it randomly} arrive into different nodes, requiring a multihop routing to reach the destination.
Wireless channels may change due to node mobility or surrounding conditions. 
Assuming the sets $\cal V$ and $\cal E$ change much slower than channel states, we fix them during the time of our interest; then a temporarily unavailable link (due to, e.g., obstacle effect and channel fading) is characterized by zero link capacity. 
Extended mobility that can lead to permanent change in network topology is not considered here.
We assume that channel states remain fixed during a timeslot, while they may change across slots. 

In wireless networks, transmission over a channel can happen only if certain constraints are imposed on transmissions over the other channels. 
An interference model specifies these restrictions on simultaneous transmissions.
We consider a family of interference models under which a node cannot transmit to more than one neighbor at the same time.
Thus, in a most general case, a node may receive packets from several neighbors while sending packets over one of its outgoing links. 
Interference constraints used by all well-known network and link layer protocols, including the general K\mbox{-}hop interference models, fall in this family.

\begin{definition}
Given an interference model, a {\it maximal schedule} is such a set of wireless channels that no two channels interfere with each other and no more channel can be added to the set without violating the model constraints.
\end{definition}
 
We describe a maximal schedule with a {\it scheduling vector} $\bs\pi\!\in\!\{0,1\}^{\vert\cal E}$ where $\pi_{ij}=1$ if channel~$ij$ is included, and $\pi_{ij}=0$ otherwise.

\begin{definition}
Given a connectivity graph $({\cal V},{\cal E})$, {\it scheduling set}~$\Pi$ is the collection of all maximal scheduling vectors. 
\end{definition}

\begin{definition}
With $\mathbb E$ denoting expectation, the {\it expected time average} of a discrete-time stochastic process $x(n)$ is defined as  
\begin{equation}\label{average1}
\overline{x} := \limsup_{\tau\to\infty}{1}/{\tau} \sum\nolimits_{n=0}^{\tau-1} \mathbb E\{x(n)\}.
\end{equation}
\end{definition}

\begin{definition}
A queuing network is {\it stable} if queue at each node $i$ and at each slot $n$, denoted as $q_i(n)$, has a bounded time average expectation, viz., $\overline{q_i} < \infty$.
\end{definition}

\begin{definition}
Given a wireless network, an arrival vector $\bs a(n)$ is {\it stabilizable} if there exists a routing policy that can make the network stable under $\bs a(n)$.
\end{definition}

For a link~$ij$, its {\it capacity} $\mu_{ij}(n)$, which is frequently called {\it transmission rate} in literature, counts the maximum number of packets the link can transmit at slot $n$.
The link {\it actual-transmission} $f_{ij}(n)$, on the other hand, counts the number of packets {\it genuinely} sent over the link at slot $n$.
Each link is also endowed with a {\it cost factor} $\rho_{ij}(n)\geqslant1$ that represents the cost of transmitting one packet over the link at slot $n$;
for example, $\rho_{ij} = \mathrm{ETX}_{ij}$, with $\mathrm{ETX}$ as defined in~\cite{Couto04}, or a cost associated with greedy embedding~\cite{Baras12}.

\subsection{Problem Statement}

For a constrained uniclass network described above, we propose HD routing policy that solves the three {\it stochastic} optimization problems as follows. 
It is important to note that these optimization problems must be solved at {\it network layer} alone, which makes it totally different from cross-layer optimization~\cite{Lin04, Neely05, Eryilmaz05, Stolyar05} that aims to control congestion by controlling arrival rates into network layer.
With no control on arrivals, the basic assumption here is that arrival rates lie within network capacity region, making the routing system stabilizable. 
Obviously, nothing prevents one to either install a flow controller on top of HD or develop an HD-based Network Utility Maximization (NUM) protocol.

{\it (Delay)}
Average network delay minimization:
\begin{equation}\label{average3}
\mathrm{Minimize}\;\;\;\, 
\overline{Q} := \sum\nolimits_{i\in\cal V}\,\overline{q_i}\;.
\end{equation}
Solving this problem for a general case requires the Markov structure of network topology process, plus arrival and channel state probabilities. 
Then in theory, the solution is obtained through dynamic programming for each possible topology along with solving a Markov decision problem. 
By even having all this required information, the number of queue backlogs and channel states increase exponentially with the size of network, which makes dynamic programming and Markov decision theory prohibitive in practice.
In fact, even for the case of a single channel, it is hard to implement the resulting stochastic algorithms~\cite{Berry02}.
While having a practical solution for a general case seems dubious, we show in Th.~\ref{minimum_queue} that HD policy solves this problem within an important class of routing algorithms, without requiring any of the above-mentioned information or dealing with any dynamic programing or Markov decision process.

{\it (Cost)}
Average quadratic routing cost minimization:
\begin{equation}\label{average2}
\mathrm{Minimize}\;\;\;\, 
\overline{R} := \!
\sum\nolimits_{ij\in\cal E} \;\overline{\!\rho_{ij}\bigl(f_{ij}\bigr){^2}}\;.
\end{equation}
The loss function $\overline{R}$, by concept, spreads out traffic with a weighted bias towards lower penalty links that reminds the optimal diffusion processes in physics, such as heat flow and electrical current \cite{Narasimhamurthi82}.
It is shown in~\cite{Georgiadis06, NeelyBook} that a stationary randomized algorithm can solve this problem.
While such an algorithm exists in theory, it is intractable in practice as it requires a full knowledge of traffic and channel state probabilities.
Further, assuming all of the probabilities could be accurately estimated, the network controller still needs to solve a dynamic programming for each topology state, where the number of states grows exponentially with the number of channels.
Nonetheless, we show in Th.~\ref{prop_fluid_model4} that HD policy solves this problem without requiring any knowledge of traffic and channel state probabilities or dealing with any dynamic programming. 

{\it (Pareto)} 
Pareto optimal performance:
\begin{equation}\label{average4}
\mathrm{Minimize}\;\;\;\, 
(1-\beta)\;\overline{Q} + \beta\,\,\overline{\!R\,}
\end{equation}
where $\beta\in[0,1]$ is a control parameter to determine relative importance between average delay and average routing cost, which naturally plays the role of {\it Lagrange multiplier} too.
To our knowledge, this is the first time in literature that such a multi-objective optimization problem is addressed in the level of solely network layer.
While even the related single-objective optimization problems are not easy to manage, we show in Th.~\ref{Pareto_performance} that within the same class of routing policies mentioned in (Cost), HD policy solves problem~\eqref{average4} subject to convex Pareto boundary on the feasible $(\overline{Q},\overline{R})$ region, with requiring no knowledge of traffic and topology.
%

\subsection{Back-Pressure (BP) Policy}

At each slot~$n$, the original BP~\cite{Tassiulas92} observes queue backlogs $q_i(n)$ at network layer and estimates channel capacities $\mu_{ij}(n)$ to make a routing decision as follows. 

\begin{enumerate}[\it 1{)}]
\item 
{\it BP weighting:} 
For every link $ij$ find link queue-differential $q_{ij}(n):=q_i(n)-q_j(n)$ and weight the link with
\begin{equation*}\label{BP2}
w_{ij}(n) := \mu_{ij}(n)\,q_{ij}(n){^+}.
\end{equation*}

\item
{\it BP scheduling:} 
Find a scheduling vector such that
\begin{equation*}\label{BP3}
\bs\pi(n) = \arg\max_{\bs\pi\in\Pi} \sum\nolimits_{ij\in\cal E}\pi_{ij}w_{ij}(n)
\end{equation*}
where ties are broken arbitrarily.

\item
{\it BP forwarding:} Over each activated link with $w_{ij}(n)>0$ transmit packets at full capacity $\mu_{ij}(n)$. 
If there is no enough packets at node~$i$, transmit null packets. 

\end{enumerate}

\subsection{V-Parameter BP Policy}

Thus far, the drift-plus-penalty approach~\cite{Georgiadis06, NeelyBook}, which we refer to as V-parameter BP hereafter, has been the only feasible approach to decreasing (not minimizing) a generic routing penalty at network layer. 
We take the V-parameter BP as a yardstick as to how HD performs.
To incorporate average routing cost $\overline R$ into the original BP, the V-parameter BP adds a usage cost to each link queue-differential via replacing the link weight of BP by 
%
\begin{equation}\label{BP4}
w_{ij}(n) := \mu_{ij}(n) \bigl( \,q_{ij}(n) - V\rho_{ij}(n)\,\mu_{ij}(n) \,\bigr){^+}
\end{equation}
where $V \in [0,\infty)$ trades queue occupancy for routing penalty, while $V\!=0$ recovers the original BP.

The V-parameter BP yields a Dirichlet routing cost within $O(1/V)$ of its minimum feasible value to the detriment of growing average delay of $O(V)$ relative to that of original BP~\cite{NeelyBook}.
Thus, the policy is not able to achieve minimum routing cost subject to finite network delay, i.e., delay grows to infinity as routing cost is pushed towards its minimum.
Another issue is that the resulting tradeoff depends on both $V$ and the network, with two negative consequences: 
(i)~The same value of $V$ leads to different levels of tradeoff in different networks, and 
(ii)~The level of tradeoff in the same network varies by topology and arrival rates, making it difficult to find a proper $V$ in practice.

\section{Heat-Diffusion (HD) Policy}
\label{s:C3_HDdesign}

To provide a convenient way of unifying our proposed scheme with the large body of previous works on BP, we design HD with the same algorithmic structure, complexity and overhead, in both computation and implementation, as BP.

\renewcommand{\tabularxcolumn}[1]{>\arraybackslash}
\definecolor{light-gray}{gray}{0.8}
\begin{table}[t!]%
\caption{Algorithmic structure of HD versus V-parameter BP in a uniclass network.
\vspace{5pt}
}
\label{table1} \centering %
\begin{tabularx}{0.75\columnwidth}{@{}l@{}c@{}c@{}c@{}}
\toprule %
\multirow{4}*{~~\begin{sideways} \colorbox{light-gray}{~Weighting~}~~ \end{sideways}}
& \multirow{2}*{~~~$\smash{\widehat{f_{ij}}}(n)$}
& ~~BP~ & $\min\bigl\{ \,\mu_{ij}(n),\: q_i(n) \bigr\}$ \\ 
\cmidrule (l){3-4}
&& ~~HD~          & $\min\bigl\{ \phi_{ij}(n)\, q_{ij}(n){^+}\! ,\: \mu_{ij}(n) \bigr\}$ \\ 
\cmidrule[0.05em] (l){2-4} 
& \multirow{2}*{~~~$w_{ij}(n)$}
& ~~BP~ & $\mu_{ij}(n) \bigl(q_{ij}(n)-V\rho_{ij}(n)\,\mu_{ij}(n) \bigr){^+}$ \\ 
\cmidrule (l){3-4}
&& ~~HD~          & ~~$2\,\phi_{ij}(n)\,
                         q_{ij}(n)\smash{\widehat{f_{ij}}}(n) - \smash{\widehat{f_{ij}}}(n)^2$ \\ 
\midrule
\multicolumn{3}{@{}l}{~~\colorbox{light-gray}{Scheduling}} & $\hspace{-5pt}\bs\pi(n) = \arg\max_{\bs\pi\in\Pi} \displaystyle\sum\nolimits_{ij\in\cal E}\pi_{ij}w_{ij}(n)$\vspace{1pt} \\ 
\midrule
\multicolumn{3}{@{}l}{~~\colorbox{light-gray}{Forwarding}} & 
$\hspace{-5pt}f_{ij}(n)=
\begin{cases}
\smash{\widehat{f_{ij}}}(n) & \text{if } \pi_{ij}(n)=1
\\[3pt]
 0 & \text{otherwise}
\end{cases}$ 
\\ 
\bottomrule
\end{tabularx}
\vspace{0pt}
\end{table}

\subsection{HD Algorithm}

At each slot~$n$, HD policy observes link queue-differentials $q_{ij}(n):=q_i(n)-q_j(n)$ at network layer and estimates channel capacities $\mu_{ij}(n)$ and channel cost factors $\rho_{ij}(n)$ to make a routing decision as follows. 

\begin{enumerate}[\it 1{)}]
\item
{\it HD weighting:} 
For every link $ij$ first calculate the number of packets it would transmit if it were activated as 
\begin{equation}\label{HD2}
\begin{gathered}
\smash{\smash{\widehat{f_{ij}}}}(n) := \min\bigl\{ \phi_{ij}(n) q_{ij}(n){^+}\! ,\: \mu_{ij}(n) \bigr\}
\\
\phi_{ij}(n) := (1\! - \!\beta)/\vartheta_{ij} + \beta/\rho_{ij}(n)
\end{gathered}
\end{equation}
where $\vartheta_{ij} = 1$ if node $j$ is the final destination, i.e., $j=d$, and $\vartheta_{ij} = 2$ otherwise.
The Lagrange control parameter $\beta$ is as defined in \eqref{average4} to make a tradeoff between queue occupancy and routing penalty, and the hat notation denotes a predicted value which would not necessarily be realized.
Then determine the link weight as 
\begin{equation}\label{HD3}
w_{ij}(n) := 2\,\phi_{ij}(n)q_{ij}(n)\smash{\smash{\widehat{f_{ij}}}}(n) - \smash{\widehat{f_{ij}}}(n)^2. 
\end{equation}
\item
{\it HD scheduling:} 
Find a scheduling vector, in the same way as BP, using the max-weight scheduling, such that
\begin{equation}\label{HD4}
\bs\pi(n) = \arg\max_{\bs\pi\in\Pi} \sum\nolimits_{ij\in\cal E} \pi_{ij}w_{ij}(n)
\end{equation}
where ties are broken arbitrarily.
\item
{\it HD forwarding:}
Over each activated link transmit $\smash{\widehat{f_{ij}}}(n)$ number of packets, viz.,
\begin{equation}\label{HD5}
{f_{ij}(n)=
\begin{cases}
\smash{\widehat{f_{ij}}}(n) & \text{if } \pi_{ij}(n)=1
\\[3pt]
 0 & \text{otherwise}
\end{cases}} 
\end{equation}
where $f_{ij}(n)$ represents the number of packets genuinely sent over link $ij$ at slot $n$.
\end{enumerate}

It is critical to discriminate among actual link transmissions $f_{ij}(n)$, link transmission predictions $\smash{\widehat{f_{ij}}}(n)$ and link capacities $\mu_{ij}(n)$.
Also notice that $\widehat{f_{ij}}(n)$ in~\eqref{HD2} could be non-integer.
In practice, the final number of packets to be transmitted over links can be rounded to the nearest integer to $\smash{\widehat{f_{ij}}}(n)$ with no important influence on the performance. 
To be more precise, however, every node may algebraically add the packet residuals sent on each of its ongoing links so as to make a compensation as soon as the sum hits either $1$ or $-1$. 

Table 1 compares HD and V-parameter BP algorithms, which emphasizes the same algorithmic structure, computational complexity and overhead signaling. 

\begin{remark} 
(i)~Since $\rho_{ij}(n)\geqslant 1$ by assumption, we get $0 < \phi_{ij}(n)\leqslant 1$ for all~$\beta\!\in\![0,1]$.
(ii)~If $q_{ij}(n) \leqslant 0$, we get $\smash{\widehat{f_{ij}}}(n)=0$ due to~\eqref{HD2} and $w_{ij}(n)=0$ from~\eqref{HD3};
in this case, even if the link were scheduled by~\eqref{HD4}, still no packet would be transmitted over it.
(iii)~If $q_{ij}(n) > 0$, we get $q_{ij}(n){^+} \!= q_{ij}(n)$ and since $\smash{\widehat{f_{ij}}}(n) \leqslant \phi_{ij}(n)q_{ij}(n)$ due to~\eqref{HD2}, the link weight~\eqref{HD3} still remains positive. 
(iv)~In light of $q_{ij}(n){^+}\!\leqslant q_i(n)$ and $\phi_{ij}(n) \leqslant 1$, the value of $\smash{\widehat{f_{ij}}}(n)$ never exceeds the number of packets in the transmitting node $i$.
\end{remark}

\begin{remark}
For $\beta=0$, HD policy reduces to the initial adiabatic-based HD policy proposed in \cite{Reza12}, where the packet forwarding follows a thermally adiabatic, and so insulated, heat process on each link.
\end{remark}

\begin{figure}[b!]   
	\centering
	\includegraphics[ scale=0.8, trim=5.5cm 5.4cm 10.5cm 7.7cm, clip=true ]{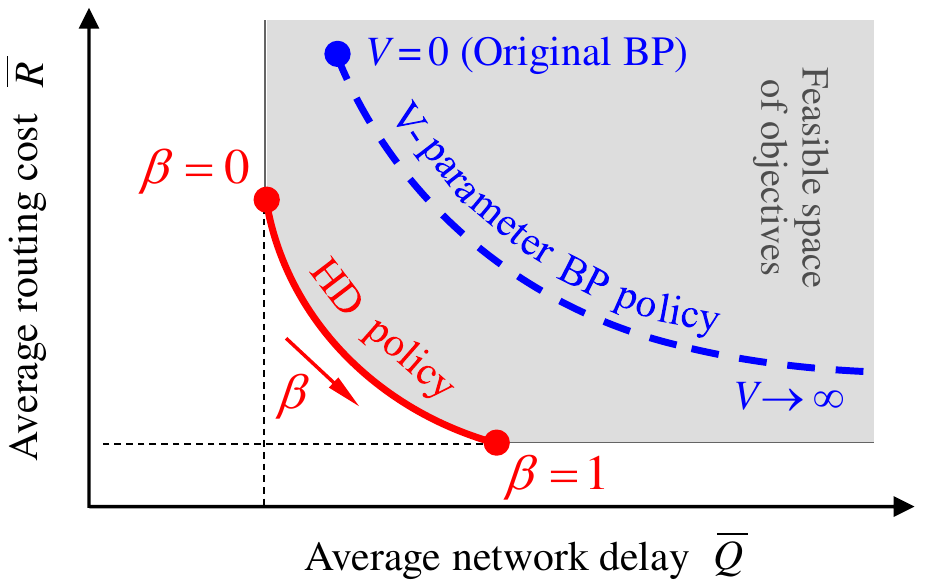}  
	\caption{
	Graphical description of HD Pareto optimality with respect to average queue congestion and the Dirichlet routing cost, compared with the performance of V-parameter BP.
	}
\label{figure1}
\end{figure}

\begin{remark}
In a special case that all links are of the same capacity, i.e., $\mu_{ij}(n)=\mu(n)$, and all link queue-differentials remain less than it, i.e., $q_{ij}(n)<\mu(n)$, HD policy with $\beta=0$ and $\alpha$-weighted policy of \cite{Shah06} with $\alpha=2$ turn to be equivalent. 
Packet switches are well suited to this special case.
It was suggested in \cite{Shah06} that a smaller $\alpha$ may lead to a lower network delay, with a non-proven conjecture that heavy traffic delay is minimized when $\alpha\to0$.  
A discussion of this was given in~\cite{Srikant09b} along with some counterexamples. 
Even if the conjecture were true, note that for a multihop routing problem, the requirement of $q_{ij}(n)\!<\mu_{ij}(n)$ would imply the network not to be in a heavy traffic condition.
\end{remark}

\subsection{Highlights of HD Design}
\label{s:C3_Highlights}

{\it H1:}
While BP is derived by link capacity $\mu_{ij}(n)$, HD emphasizes on actual number of transmittable packets $\smash{\widehat{f_{ij}}}(n)$, though it also implicitly takes the link capacity into account through~\eqref{HD2}. 
Thus, HD allocates resources based only on genuinely transmittable packets, without counting on null packets as being practiced in BP schemes. 

{\it H2:}
The link weight~\eqref{HD3}, which itself directly controls the scheduling optimization problem, is taken quadratic in the link queue-differential $q_{ij}(n)$, where for $\phi_{ij}(n)q_{ij}(n)\leqslant\mu_{ij}(n)$ is simplified to $w_{ij}(n)=\phi_{ij}(n){^2}q_{ij}(n){^2}$.
This contrasts with BP~weighting $w_{ij}(n)=\mu_{ij}(n)q_{ij}(n)$ which is linear in $q_{ij}(n)$.
The quadratic weight is central to HD key property (Th.~\ref{key_property}) which is fundamental to other HD qualities.  
 
{\it H3:}
Varying the penalty factor $\beta$ makes a {\it universal} tradeoff in performance that depends neither on network nor on arrivals with the following significant results:
\setdefaultleftmargin{1.25em}{}{}{}{}{} 
\begin{compactitem}
\item
HD is throughput optimal for all $\beta\in[0,1]$ (Th.~\ref{maximum_throughput}).
\item 
At $\beta=0$, the average total queue $\overline{Q}$, and so average network delay, decrease to their minimum feasible values within the class of routing policies that rely only on present queue backlogs and current channel states (Th.~\ref{minimum_queue}). 
\item 
Raising $\beta$ adds to average delay in return for a lower routing cost, where the exclusive merit of HD is to provide the best tradeoff between these two criteria (Th.~\ref{Pareto_performance}).
\item
At $\beta=1$, the average routing cost $\overline{R}$ reaches its minimum (Th.~\ref{prop_fluid_model4}) through an optimal tradeoff with average network delay.
Note that in V-parameter BP, network delay grows to infinity as routing cost is pushed towards its minimum.
\end{compactitem}

{\it H4:}
Unlike BP that forwards the highest possible number of packets over activated links, HD controls packet forwarding by limiting it to $\phi_{ij}(n)q_{ij}(n)$ with $\phi_{ij}$ changing between 0 and 1 as a function of $\beta$, $\vartheta_{ij}$ and $\rho_{ij}$. 
This reduces {\it queue oscillations} by decreasing unnecessary packet forwarding across links, which itself reduces total power consumption and routing penalty.
Thus, it is not surprising to see that $\phi_{ij}$ is decreasing, and so as to have a higher impact, by increasing $\beta$ that means more emphasis on routing penalty.
Forwarding a portion of link queue-differentials rather than filling up link capacities also complies with resembling {\it heat flow} on the underlying directed graph (Th.~\ref{prop_fluid_model2}) that in effect minimizes time average routing cost in light of Dirichlet's principle (Th.~\ref{prop_fluid_model4}).

Figure~\ref{figure1} provides a graphical comparison between operation of HD for ${\beta\in[0,1]}$ and V-parameter BP for $V\in[0,\infty)$.
The performance region is restricted to the set of all $\overline Q$ achievable by the class of all routing policies that act based only on present queue backlogs and current channel states, and is assumed to have a convex Pareto boundary.

\subsection{Illustrative Examples}

In order to focus merely on the policy itself, we take everything deterministic in our examples here, resting assure that the results purely show the policy performance not contaminated by stochastic effects.
We however know that all HD properties are analytically proven for stochastic arrivals and random topologies under very general conditions.

\begin{figure}[t!]
	\centering
	\includegraphics[ scale=.8, trim=5.5cm 5.1cm 5.8cm 7.5cm, clip=true ]{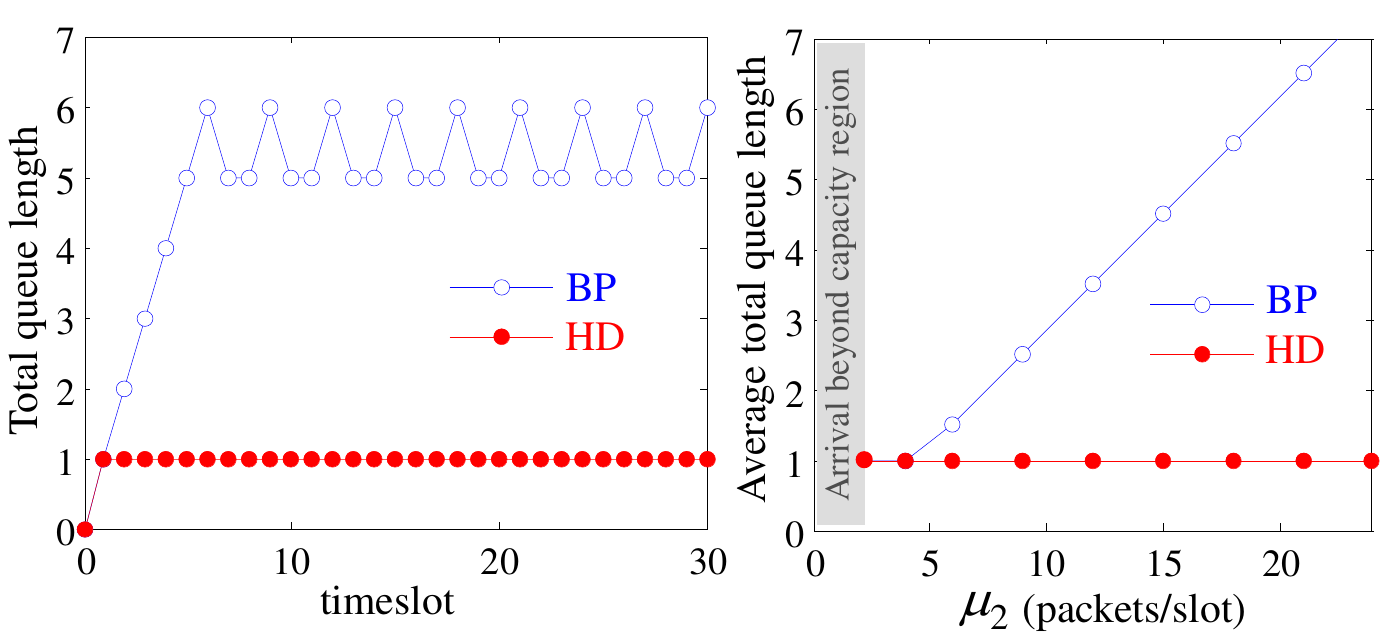}  
	\vspace{-6pt}
	\caption{
	Two-queue downlink: Performance of HD with $\beta=0$ versus original BP.
	While for all admissible link capacities total queue is minimized under HD, it grows linearly in $\mu_2$ under BP.
	}
\label{figure2}
\end{figure}

\begin{figure}[b!]
	\centering
	\includegraphics[ scale=.8, trim=5.5cm 5.3cm 5.8cm 7.7cm, clip=true ]{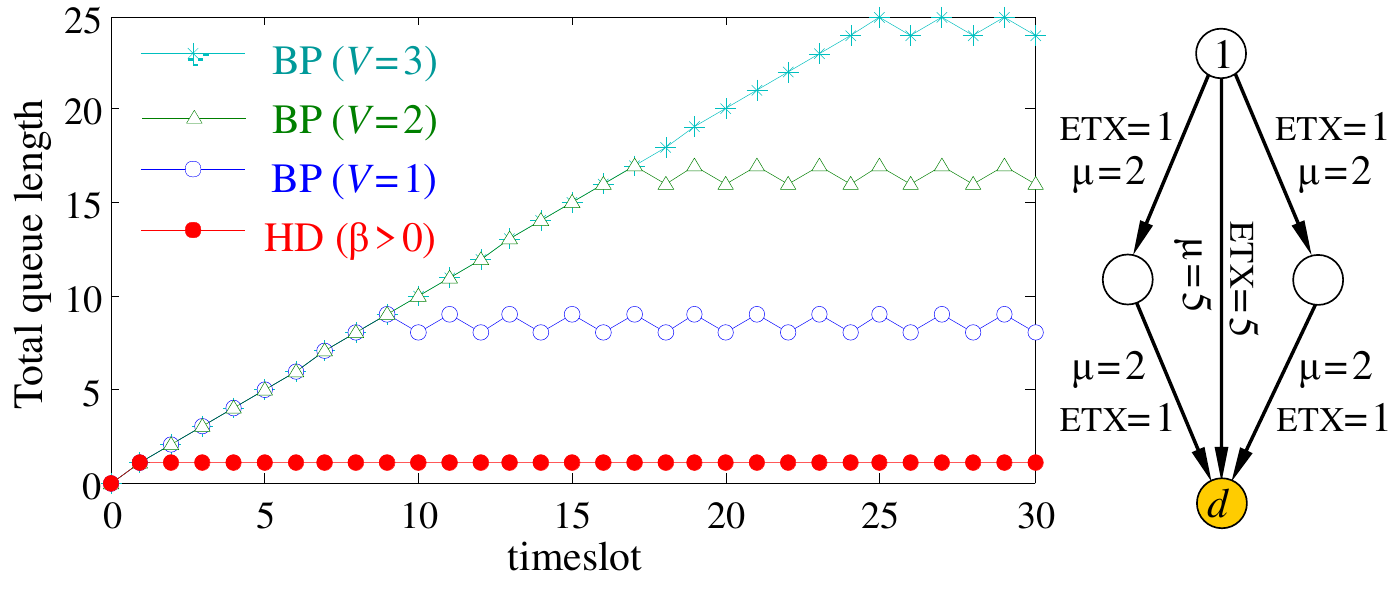}  
	\vspace{-6pt}
	\caption{
	Lossy link network: Performance of HD versus V-parameter BP. 
	While total queued packets is stabilized at 1 under HD for any $\beta>\nobreak 0$, it grows linearly in $V$ under V-parameter BP. 
	}
\label{figure3}
\end{figure}

{\it Two-queue downlink:}
Consider a base station that transmits data to two downlink users, where at most one link can be activated at each timeslot. 
Let link~1 be of constant capacity $\mu_1 = 3$ (packets/slot) and link 2 of time-varying capacity $\mu_2 \geqslant 2$.
Assume one packet to arrive for each user at every timeslot.
It is then easy to verify that for $\mu_2 < 1.5$, the given arrival goes beyond the network capacity region.

For $q_1(0)=q_2(0)=0$, Fig.~\ref{figure2} compares the performance of HD with $\beta=0$ and original BP.
The left side panel depicts timeslot evolution of $q_1(n)+q_2(n)$ for $\mu_2=18$. 
The right side panel shows the steady-state average of total queue length as a function of $\mu_2$. 
For $2 \leqslant \mu_2 \leqslant 5$, both HD and BP perform the same.
For $\mu_2 \geqslant 5$, however, average total queue length increases linearly in $\mu_2$ under BP, while HD holds the optimal performance for all admissible link capacities.
This exemplifies H1 in the previous subsection, i.e., the efficiency of link scheduling based on actual transmittable packets rather than link capacities.

{\it Lossy link network:}
Consider the 4-node network of Fig.~\ref{figure3} with lossy links and subject to 1\mbox{-}hop interference model, i.e., two links with a common node cannot be activated at the same time.
The links are labeled with both ETX and capacity, where ETX is a quality metric defined as the expected number of data transmissions required to send a packet without error over a link~\cite{Couto04}.
Assume that at every timeslot a single packet arrives at node~1 destined for node~$d$.
Following~\cite{Moeller10}, let us take $\rho_{ij}=\nobreak\mathrm{ETX}_{ij}$.

\begin{figure}[t!]
	\centering
	\hspace{-0pt}
	\includegraphics[ scale=.8, trim=5.5cm 5.9cm 5.8cm 7.5cm, clip=true ]{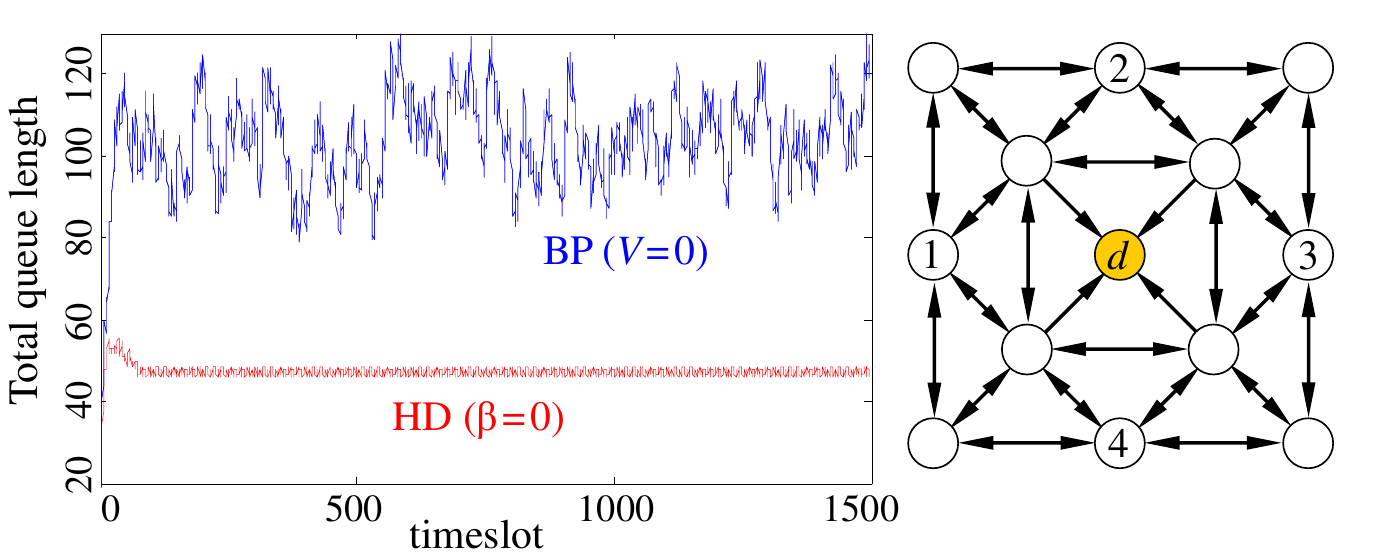}  
	\vspace{-4pt}
	\caption{
	Power minimization: Timeslot evolution of total queue backlog in HD with $\beta=0$ versus original BP, showing the minimization of average queue congestion by HD. 
	Noticeable is also the little steady-state oscillations in total queue under HD contrary to its large variations under BP.
	}
\label{figure4}
\end{figure}

\begin{figure}[b!]
	\centering
	\includegraphics[ scale=.8, trim=5.6cm 5.5cm 5.6cm 7.7cm, clip=true ]{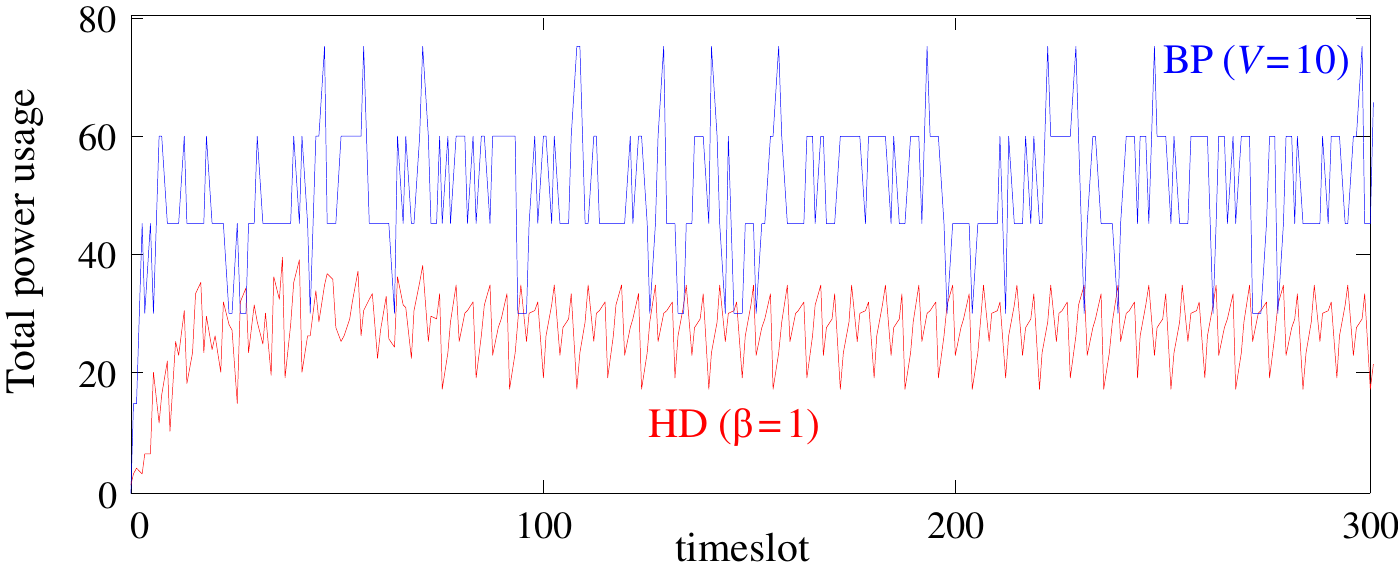}  
	\vspace{-6pt}
	\caption{
	Power minimization: Timeslot evolution of total power consumption, which is highly correlated with the Dirichlet routing cost, in HD with $\beta=1$ versus V-parameter BP with $V=10$. 
	}
\label{figure6}
\end{figure}

For zero initial conditions, Fig.~\ref{figure3} compares the performance of HD with V-parameter BP.
While HD easily stabilizes total queued packets at 1 for any $\beta>\nobreak 0$, trying with different values of $V$ indicates the weakness of V-parameter BP in aptly supporting the arrival. 
This simplistically shows one of the impacts of entering link cost factor~$\rho_{ij}$ as a {\it multiplicand} in the HD weighting formula~\eqref{HD3} rather than an {\it addend} in the V-parameter BP weighting formula~\eqref{BP4}. 

{\it Power minimization:}
Consider the sensor network of Fig.~\ref{figure4} subject to 1\mbox{-}hop interference model. 
Suppose that each link $ij$ has a noise intensity $N_{ij}\in[1,5]$ which is randomly assigned at first and keeps fixed during the simulation.
For each link, we adopt Shannon capacity $\mu_{ij}=\Omega_{ij}\log_2(1+P_{ij}/N_{ij})$ with $P_{ij}$ as power transmission and $\Omega_{ij}$ as bandwidth.
At every timeslot, two packets arrive at nodes 1, 2, 3 and 4, destined for node~$d$.
The aim is to minimize total $\smash{\rho_{ij}(f_{ij}){^2}}$ with $\rho_{ij}:=P_{ij}/\mu_{ij}$, which implicitly minimizes total power consumption in the network.
For simplicity, let us fix $P_{ij}=15$ and $\Omega_{ij}=5$ for all links so that the capacity on each link is decided only by its noise intensity.

\begin{figure}[t!]
	\centering
	\hspace{-0pt}
	\includegraphics[ scale=.8, trim=5.5cm 4.3cm 5cm 7.7cm, clip=true ]{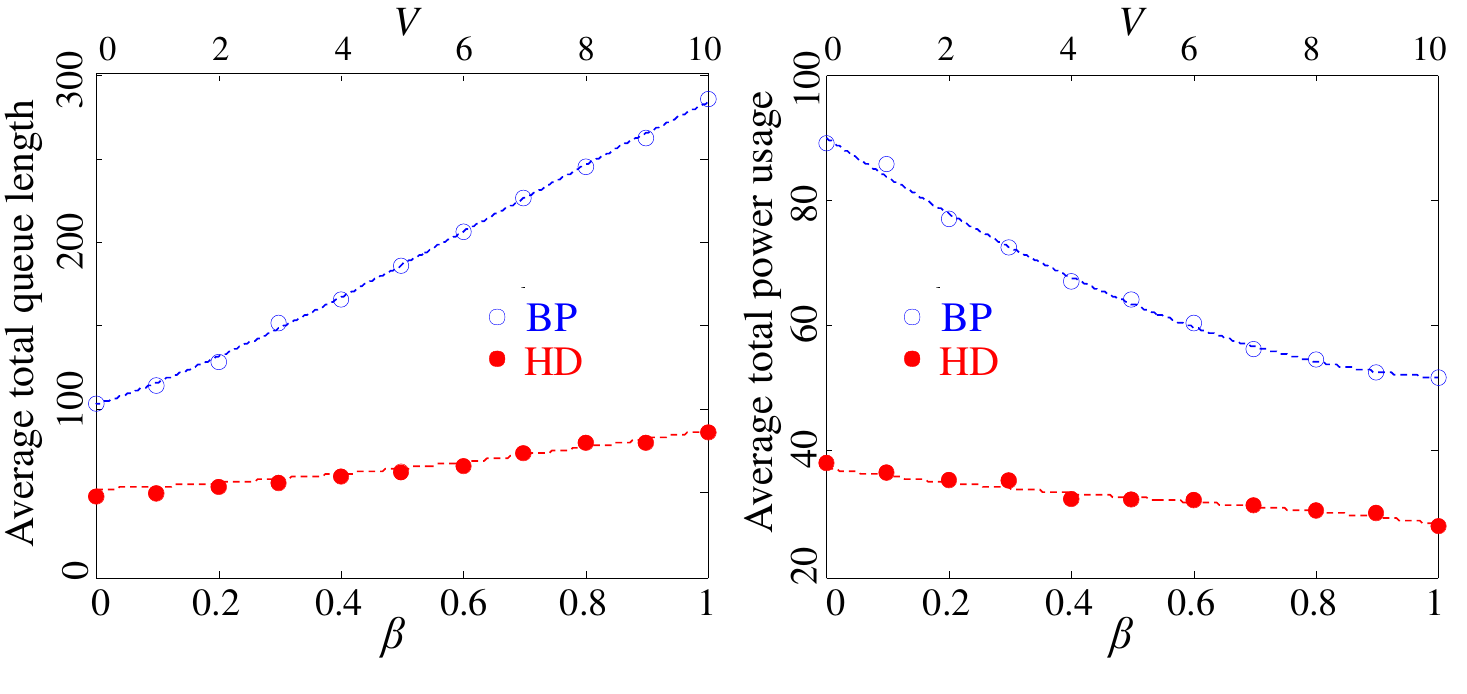}  
	\vspace{-10pt}
	\caption{
	Power minimization: Trading queue congestion for power consumption by HD as a function of $\beta$ and by V-parameter BP as a function of $V$, with the dashed lines representing interpolation.
	}
\label{figure5}
\end{figure}

Figure~\ref{figure4} displays timeslot evolution of total queue length for HD with $\beta=0$ and for the original BP ($V\!=0$).
Average queue congestion is minimized at about 50 packets under HD, compared with over 100 packets under original BP. 
Further, little steady-state oscillations in total queue congestion under HD contrary to its large variations under BP verifies H4 in the previous subsection.

In minimizing average routing cost, Fig.~\ref{figure6} displays timeslot evolution of total power consumption for HD with $\beta=1$ and for V-parameter BP with $V=10$. 
Note that while the total power consumption and the average routing cost are not identical, they are highly correlated with each other. 
Smaller steady-state oscillations in total power under HD endorses both H1 and H4 in the previous subsection, showing the defect of link capacity-driven scheduling and maximum packet forwarding by BP. 

Figure~\ref{figure5} displays the tradeoff between queue congestion and power usage in HD as a function of $\beta$ and in V-parameter BP as a function of $V$.
The results verify H3 in the previous subsection and concur with the graphical illustration of HD Pareto optimal performance depicted by Fig.~\ref{figure1}.
They also match the timeslot evolution results displayed in Fig.~\ref{figure4} for total queue length at $\beta=0$ and $V=0$, and in Fig.~\ref{figure6} for total power consumption at $\beta=1$ and $V=10$. 
Note the rapid growth of queue lengths in V-parameter BP when average power usage is pushed downwards, indicating the fact that the V-parameter BP cannot reach the minimum routing cost subject to network stability, i.e., bounded queue lengths.

\section{Key Property of HD Policy}
\label{s:C3_key}

%

Consider a general uniclass queuing network with a single destination node $d$.
As before, let $q_i(n)$ be the number of existing packets at node $i$ at slot~$n$. 
State variables of the system can then be represented by the following vector: 
\begin{equation*}\label{model1}
\bs q_\circ(n):=\bigl[\,q_1(n), \ldots, q_{d-1}(n), q_{d+1}(n), \ldots, q_{\vert\cal V\vert}(n)\,\bigr].
\end{equation*}
Note that $q_d(n) \equiv 0$ is discarded from state variables.

\begin{notation}
We use subscript $\circ$ to denote reduced vectors or matrices obtained by discarding the entries corresponding to the destination node $d$.
\end{notation}

Let a stochastic process $a_i(n)$ represent the number of exogenous packets arriving into node $i$ at slot~$n$.
Discard $a_d(n) \equiv 0$ and compose the vector of node arrivals as
\begin{equation*}\label{model2}
\bs a_\circ(n):=\bigl[\,a_1(n), \ldots, a_{d-1}(n), a_{d+1}(n), \ldots, a_{\vert\cal V\vert}(n)\,\bigr].
\end{equation*}
Also compose the vector of link actual transmissions as
\begin{equation*}\label{model3}
\bs f(n):=\bigl[\,f_1(n), \ldots, f_{\vert\cal E\vert}(n)\,\bigr]
\end{equation*}
where, as before, $f_{ij}(n)$ represents the number of packets actually sent over link $ij$ at slot $n$.

Given a directed graph $(\cal V,\cal E)$, let $\bs B$ denote the {\it node-edge} incidence matrix, in which $B_{i\ell}$ is 1 if node~$i$ is the tail of {\it directed} edge~$\ell$, $-1$ if $i$ is the head, and 0 otherwise.\footnote{
In combinatorial geometry, one can view graph as a 1-complex, where $\bs B$ is its 1-incidence matrix that describes the correlation between all oriented 1-cells (edges) and 0-cells (nodes) in the complex.}
Then $\bs B_\circ$ denotes a reduction of $\bs B$ that discards the row related to the destination node $d$, which is referred to as reduced incidence matrix. 
One can verify that $\bs B_\circ\bs f(n)$ is a node vector, in which the entry corresponding to node $i$ reads the net outflow as
\begin{equation*}
(\bs B_\circ\bs f)_i(n)=\sum\nolimits_{b\in\out(i)}f_{ib}(n)-\sum\nolimits_{a\in\into(i)}f_{ai}(n)
\end{equation*}

Using the above notation, the $\bs f$-controlled, stochastic state dynamics of a uniclass queuing network is captured by 
\begin{equation}\label{model5}
\bs q_\circ(n+1)=\bs q_\circ(n) + \bs a_\circ(n) - \bs B_\circ\bs f(n)\,. 
\end{equation}

Note that the link capacities $\mu_{ij}(n)$ vary by channel states, while the link actual transmissions $\bs f_{ij}(n)$ are assigned by a routing policy subject to 
$0 \leqslant f_{ij}(n) \leqslant \min\{q_i(n),\,\mu_{ij}(n)\}$.
This difference explains why despite traditional notation in literature, there is no need for $(\cdot){^+}$ operation in the queue equation~\eqref{model5}.

In the wake of \eqref{model5}, the next theorem formalizes the HD main characteristic, which is central to the proof of Th.~\ref{maximum_throughput} on HD throughput optimality, Th.~\ref{minimum_queue} on HD average network delay minimization, Th.~\ref{prop_fluid_model2} on connection between HD fluid limit and combinatorial heat equation, and Th.~\ref{prop_fluid_model4} on HD average quadratic routing cost minimization.
Before proceeding to the theorem, let us define the link weight matrix as
\begin{equation}\label{phi}
\bs\Phi(n):=\diag(\bs\phi(n)) 
\end{equation}
where $\bs\phi(n)$ represents the vector composed of $\phi_{ij}(n)$ as defined in \eqref{HD2}.

\begin{theorem}[HD Key Property]
\label{key_property}
Consider a uniclass wireless network constraint by capacity, directionality and interference. 
At every timeslot~$n$ and for all $\beta\in[0,1]$, HD policy maximizes the $\bs f$-controlled functional
\begin{equation}\label{Dfunction2}
D(\bs f,\bs q_\circ,n) := 2\,\bs f(n)^{\!\top}\bs\Phi(n)\bs B\otop\bs q_\circ(n) - \bs f(n)^{\!\top}\bs f(n). 
\end{equation}
\end{theorem}

Consider the long-term average of functional $D(\bs f,\bs q_\circ,n)$ defined as
\begin{equation}\label{Dfunction2a}
\overline{D}(\,\overline{\bs f},\,\overline{\bs q_\circ\!}\,) := 2\;\overline{\bs f}^\top \overline{\bs\Phi}\,\bs B\otop\,\overline{\bs q_\circ\!}\, \,-\, \overline{\bs f}^\top \overline{\bs f}\,. 
\end{equation}
Next assumption is being used in the analytical proofs of HD properties, stating that the greedy maximization of $D(\bs f,\bs q_\circ,n)$ at each timeslot leads to its maximum long-term average.
The assumption implies that one can apply the Bellman's principle of optimality, and so dynamic programming, to maximize $\overline{D}$.
It also implicitly means no overlapping among slot-based substructures of $\overline{D}$ maximization problem.

\begin{assumption}\label{ass2}
Consider a uniclass wireless network constraint by capacity, directionality and interference. 
Given a combination of network topology and traffic rates, timeslot maximization of $D(\bs f,\bs q_\circ,n)$ is an optimal substructure for global maximization of $\overline{D}(\,\overline{\bs f},\,\overline{\bs q_\circ\!}\,)$.
\end{assumption}
 
In practice, almost every wireless mesh network meets this assumption.
As an example that fails the requirement though, consider the case where exogenous packets arrive only to one node, say $a$, which is connected directly to the final destination.
Assume that all links are bidirectional with unit cost factors and infinite capacities, and so link interference is the only network constraint. 
Obviously, depleting the whole queue into the destination maximizes $D$ to $q_a(n)^2$ at each timeslot.
To maximize $\overline{D}$, however, a portion of traffic must be forwarded through other paths that connect node $a$ to the destination.

\section{HD Throughput Optimality}
\label{s:C3_throughput}

Let the stochastic process $\bs S(n)=\bigl( S_1(n),\cdots,S_{\vert{\cal E}\vert}(n) \bigr)$ represent channel states at slot~$n$, describing all uncontrollable factors that affect wireless link capacities and cost factors. 
We assume that $\bs S(n)$ evolves according to an ergodic stationary process and takes values in a finite set $\cal S$.
Thus, by Birkhoff's ergodic theorem, each state $\bs S \in \cal{S}$ has a probability of
\begin{equation}\label{topology0}
s:=\mathbb P\bigl\{\bs S(n)\!=\!\bs S\bigr\}=
\limsup_{\tau\to\infty} {1}/{\tau} \sum\nolimits_{n=0}^{\tau-1} {\mathbb I}_{\bs S(n)=\bs S} 
\end{equation}
where $\sum_{\bs S \in \cal S} s = 1$.
Then the expected link capacities and cost factors are obtained as
\begin{align} 
\mathbb E \bigl\{\bs\mu(n)\}  = & 
\sum\nolimits_{\bs S \in \cal S} s \: \mathbb E \bigl\{\bs\mu(n) 
\bigl\vert \bs S(n) = \bs S \bigr\}
\label{topology1}
\\
\mathbb E \bigl\{\bs\rho(n)\} = & 
\sum\nolimits_{\bs S \in \cal S} s \: \mathbb E \bigl\{\bs\rho(n)  
\bigl\vert \bs S(n) = \bs S \bigr\}
\label{topology2}
\end{align} 
where $\bs\mu(n)$ and $\bs\rho(n)$ represent the vectors composed of link capacities $\mu_{ij}(n)$ and link cost factors $\rho_{ij}(n)$, respectively. 

Note that the existence of probability distribution \eqref{topology0} or expected values \eqref{topology1} and \eqref{topology2} by no means imply that they are known to a routing policy.
Specifically, HD performs without knowing any of these information. 
Nonetheless, the ergodicity of $\bs S(n)$ along with the law of large numbers imply
\begin{align*} 
\mathbb E \bigl\{\bs\mu(n)\}  = & 
\lim_{\tau\to\infty}{1}/{\tau} \sum\nolimits_{n=0}^{\tau-1} \bs\mu(n)
\\
\mathbb E \bigl\{\bs\rho(n)\} = & 
\lim_{\tau\to\infty}{1}/{\tau} \sum\nolimits_{n=0}^{\tau-1} \bs\rho(n)
\end{align*} 
meaning that the expectations converge to the long-term averages.
Thus, a routing policy could estimate $\mathbb E\{\bs\mu(n)\}$ and $\mathbb E\{\bs\rho(n)\}$ by observing timeslot variables $\bs\mu(n)$ and $\bs\rho(n)$ for a long enough period of time, at least in theory. 
This justifies the existence of stationary randomized policies that base their routing decisions only on arrival statistics and channel state probabilities, but fully independent of queue occupancies.

\subsection{Characteristic of Network Capacity Region}
\label{s:C3_channel}

Consider a uniclass wireless network that is described by a connectivity graph $(\cal{V},\cal{E})$, a destination node $d$, and an ergodic stationary channel state process~$\bs S(n)$.

\begin{definition}
Given a routing policy, its {\it stability region} is the set of all arrival vectors that it can stably support, i.e., make the network stable under those arrivals. 
\end{definition}

\begin{definition}
Given a network layer, its {\it capacity region} is the union of stability regions achieved by all routing policies, including those which are possibly unfeasible. 
\end{definition}

It can be shown that for any network, its capacity region is convex and compact and so is closed and bounded \cite{Georgiadis06}. 

\begin{definition}
A routing policy is {\it throughput optimal} if it can stabilize the entire network capacity region, i.e., secure queue stability under all stabilizable arrival vectors.
\end{definition}

An arrival vector $\bs a_\circ(n)$ is in the network capacity region, i.e., {\it stabilizable,} if and only if there exists a set of link actual transmissions $\bs f(n)$ that satisfy
\begin{equation}\label{hyperflow1}
\overline{a_i} = \sum\nolimits_{b\in \out(i)}\overline{f_{ib}} \,- \sum\nolimits_{a\in \into(i)}\overline{f_{ai}} \,\,,
\;\, \forall \, i\in{\cal V}\setminus\!\{d\}
\end{equation}
constrained by link capacities and interference.
Under an ergodic channel state process, this basically reads the long-term average flow conservation at the nodes. 
In a matrix form, \eqref{hyperflow1} can equivalently be shown by $\overline{\bs a_\circ\!}\,=\bs B_\circ\overline{\bs f}$.

\begin{remark}
Link actual transmissions $\bs f(n)$ are not fixed, but depend on routing policy. 
Further, there could potentially exist infinite number of routing policies that meet \eqref{hyperflow1} for any stabilizable $\bs a_\circ(n)$.
Among them are the ones that use the simple probability concept of distributing packets randomly so that the desired time averages~\eqref{hyperflow1} can be achieved.
These stationary randomized policies are prohibitive in practice as they typically require perfect knowledge of arrival statistics and channel state probabilities along with an expensive computation. 
Nonetheless, the fact that these {\it queue-independent} policies exist plays a crucial role in the analytical proof of HD properties in this and next section.
\end{remark}

\subsection{HD Throughput Optimality for all $\beta$}

To prove network stability under HD policy, as well as some other HD properties in next sections, we are compelled to choose unorthodox Lyapunov candidates based on the following nonsymmetric system matrix: 
\begin{equation}\label{Lyapunov1}
\bs M_{\!\circ}(n):=\bigl(\bs B_\circ\bs B\otop\bigr){^{-1}}\bs B_\circ\bs\Phi(n)\bs B\otop.
\end{equation}
Handling Lyapunov arguments turns to be a lot more challenging, since the easy way of working with symmetric positive definite matrices ceases to exist here. 
Nonetheless, the specific structure of $\bs M_{\!\circ}(n)$ makes the following lemmas possible.

\begin{lemma}\label{Lemma1}
Given a connected uniclass wireless network, $\bs M_{\!\circ}(n)$ is {\it pseudo positive definite} in the sense that all of its eigenvalues are positive and $\bs x^{\!\top}\bs M_{\!\circ}(n)\,\bs x\geqslant 0$ for any vector $\bs x\in{\mathbb R}^{\vert{\cal V}\vert-1}$, with equality if and only if $\bs x = \bs 0$. 
\end{lemma}

\begin{lemma}\label{Lemma2}
Given a connected uniclass wireless network, for any vector $\bs x\in{\mathbb R}^{\vert{\cal V}\vert-1}$, the following identity holds:
\begin{equation}\label{Lem1}
\bs B\otop \bs M_{\!\circ}(n)\, \bs x = \bs\Phi(n)\bs B\otop \bs x\,.
\end{equation}
\end{lemma}

\begin{lemma}\label{Lemma3}
Given a connected uniclass wireless network, there exists such a scalar $1\leqslant\eta\leqslant3$ that for any vectors $\bs x,\bs y \in{\mathbb R}^{\vert{\cal V}\vert-1}$, the following inequality holds:
\begin{equation}\label{Lem2}
\bs x^{\!\top} \big( \bs M_{\!\circ}(n)^{\!\top} \!+ \bs M_{\!\circ}(n) \big) \,\bs y
\leqslant
\eta\;\bs x^{\!\top} \bs M_{\!\circ}(n) \,\bs y\,. 
\end{equation}
\end{lemma}

To analyze the HD throughput optimality, consider the Lyapunov candidate 
\[
W(n) := \bs q_{\circ\!}(n)^{\!\top\!} \bs M_{\!\circ}(n) \bs q_{\circ\!}(n).
\]
Though $W(n)$ is indeed an energy function in light of Lem.~\ref{Lemma1}, due to the nonsymmetric weighting matrix $\bs M_{\!\circ}(n)$, it has no trivial interpretation of a specific energy in the system.
Nonetheless, it clearly penalizes high queue differentials across links, compelling a more even distribution of packets over the network. 
It also incites transmission over the links of lower cost factors, leading to a less expensive routing decision.
Note that either at $\beta\!=\!0$ or for the case that all links are of the same cost factor, $\bs\Phi(n)$ is simplified to a scaled identity matrix that leads to $\bs M_{\!\circ}(n)=\bs\Phi(n)$, which in turn reduces $W(n)$ to the sum of squares of queue lengths -- a familiar Lyapunov function in most of previous results in literature. 

Let $\Delta W(n):=W(n+1)-W(n)$ be the Lyapunov drift.
Substituting for $\bs q_\circ(n+1)$ from~\eqref{model5} leads to 
\begin{equation*}
\begin{aligned}
\Delta W(n) = 
\bigl(&\bs a_\circ(n) - \bs B_\circ\bs f(n) \bigr)^{\!\top}
\bigl(\bs M_{\!\circ}(n) + \bs M_{\!\circ}(n)^{\!\top}\bigr) \bs q_\circ(n) 
\\ &
+\bigl(\bs a_\circ(n) - \bs B_\circ\bs f(n) \bigr)^{\!\top} \bs M_{\!\circ}(n) \bigl(\bs a_\circ(n) - \bs B_\circ\bs f(n) \bigr).
\end{aligned}
\end{equation*}
Let us drop timeslot variable $(n)$ for ease of notation.
Applying Lem.~\ref{Lemma3} to the first line of the above drift equation yields
\begin{equation*}\label{Lyapunov3}
\Delta W \leqslant 
\eta\,(\bs a_\circ \!-\! \bs B_\circ\bs f )^{\!\top}
\bs M_{\!\circ} \bs q_\circ 
+ (\bs a_\circ \!-\! \bs B_\circ\bs f )^{\!\top} \bs M_{\!\circ} (\bs a_\circ \!-\! \bs B_\circ\bs f )
\end{equation*}
with $1\leqslant\eta\leqslant3$. 
Let us replace $\bs f^\top\! \bs B\otop\bs M_{\!\circ}\bs q_\circ$ by $\bs f^\top\! \bs\Phi\bs B\otop\bs q_\circ$ in light of Lem.~\ref{Lemma2}, 
add and subtract the term $\frac{1}{2}\,\eta\,\bs f^\top\!\bs f$, 
and use the $D(\bs f,\bs q_\circ,n)$ expression in \eqref{Dfunction2} to obtain 
\begin{equation*}\label{Lyapunov4}
\begin{aligned}
\Delta W \leqslant \eta\, \bs a\otop \bs M_{\!\circ}\bs q_\circ  & 
- \frac{\eta}{2}\,D(\bs f,\bs q_\circ,n) - \frac{\eta}{2}\,\bs f^{\!\top}\!\bs f
\\ &
+ (\bs a_\circ - \bs B_\circ\bs f )^{\!\top} \bs M_{\!\circ} (\bs a_\circ - \bs B_\circ\bs f ).
\end{aligned}
\end{equation*}
Taking conditional expectation from the latter given the current queue backlogs $\bs q_\circ(n)$ and knowing that the term $\eta\,\bs f^{\!\top}\!\bs f$ has a zero lower bound lead to 
\begin{gather}
\mathbb E\bigl\{\Delta W\vert\bs q_\circ\bigr\} \leqslant 
\;\eta\,\mathbb E\bigl\{\bs a\otop\bs M_{\!\circ}\bigl\vert\bs q_\circ\bigr\}\,\bs q_\circ 
- \frac{\eta}{2}\,\mathbb E\bigl\{D(\bs f,\bs q_\circ)\bigl\vert\bs q_\circ\bigr\}
\nonumber
\\ 
+\, \mathbb E\bigl\{
(\bs a_\circ \!-\! \bs B_\circ\bs f )^{\!\top} \bs M_{\!\circ} (\bs a_\circ \!-\! \bs B_\circ\bs f )
\,\bigl\vert\bs q_\circ\bigr\} 
\label{th2_a}
\end{gather}
where the conditional expectation is with respect to the randomness of arrivals, channel states and routing decision -- in case of a randomized routing algorithm.

Observe that $\bs M_{\!\circ}(n) = \bigl(\bs B_\circ\bs B\otop\bigr){^{-1}}\bs B_\circ\bs\Phi(n)\bs B\otop$ is a function only of control parameter $\beta$ and link cost factors $\rho_{ij}(n)$. 
Since arrivals are independent of both $\beta$ and $\rho_{ij}$, we get 
\[
\mathbb E\bigl\{\bs a\otop\bs M_{\!\circ}\bigl\vert\bs q_\circ\bigr\} = 
\mathbb E\bigl\{\bs a\otop\bigl\vert\bs q_\circ\bigr\}
\, \mathbb E\bigl\{\bs M_{\!\circ}\bigl\vert\bs q_\circ\bigr\}.
\]
At the same time, both $\beta$ and $\rho_{ij}$ are independent of $\bs q_\circ$, so is $\bs M_{\!\circ}$, which means
$\mathbb E\{\bs M_{\!\circ}\vert\bs q_\circ\} = \mathbb E\{\bs M_{\!\circ}\}$.
On the other hand, since the network layer routing controller has no impact on arrivals, $\bs a_\circ(n)$ turns to be an independent system variable that is not influenced by anything, which implies $\mathbb E\{\bs a\otop\vert\bs q_\circ\} = \mathbb E\{\bs a\otop\}$.
Putting these results together yields
\begin{equation}\label{th2_b}
\mathbb E\bigl\{\bs a\otop\bs M_{\!\circ}\bigl\vert\bs q_\circ\bigr\}\,\bs q_\circ = 
\mathbb E\{\bs a\otop\} \, \mathbb E\{\bs M_{\!\circ}\} \,\bs q_\circ.
\end{equation}

Given the current queue backlogs $\bs q_\circ(n)$, let $\bs f^\star(n)$ be the link actual transmissions provided by HD policy.
As compared to any alternative transmission decision $\bs f(n)$, Th.~\ref{key_property} secures $D(\bs f^\star\!,\bs q_\circ,n)\geqslant D(\bs f,\bs q_\circ,n)$ for all $\beta$ and at each slot $n$.
Considering this with the equality~\eqref{Lem1} of Lem.~\ref{Lemma2} implies 
\begin{equation*}
D(\bs f^\star\!,\bs q_\circ,n) \geqslant 2\,\bs f^\top\!\bs B\otop\bs M_{\!\circ}\,\bs q_\circ - \bs f^\top\!\bs f.
\end{equation*}
Taking conditional expectation given current queues yields
\begin{equation*}
\mathbb E\bigl\{D(\bs f^\star\!,\bs q_\circ,n)\bigl\vert\bs q_\circ\bigr\} \geqslant 
2\,\mathbb E\bigl\{\bs f^\top\!\bs B\otop\bs M_{\!\circ}\bigl\vert\bs q_\circ\bigr\}\,\bs q_\circ 
- \mathbb E\bigl\{\bs f^\top\!\bs f\,\bigl\vert\bs q_\circ\bigr\}.
\end{equation*}

As one alternative transmission decision $\bs f(n)$ to be compared with the $\bs f^\star(n)$ provided by HD policy, consider the case where $\bs f(n)$ is produced by a routing algorithm which makes independent, stationary and randomized transmission decisions at each slot~$n$ based only on arrivals and link capacities and so independent of both queue backlogs and link cost factors~\cite{Georgiadis06}.
Let us fix $\bs f(n)$ for such an algorithm and refer to it as $\bs{f^\prime}(n)$. 
Using equality $\mathbb E\{\bs M_{\!\circ}\vert\bs q_\circ\} = \mathbb E\{\bs M_{\!\circ}\}$ and considering that $\bs{f^\prime}(n)$ is independent from $\bs q_\circ(n)$ and $\bs M_{\!\circ}(n)$, we obtain 
\begin{equation*}
\mathbb E\bigl\{D(\bs f^\star\!,\bs q_\circ,n)\bigl\vert\bs q_\circ\bigr\} \geqslant 
2\,\mathbb E\{\bs{f^\prime}{^\top\!}\bs B\otop\} \,
\mathbb E\{\bs M_{\!\circ}\}\,\bs q_\circ 
- \mathbb E\{\bs{f^\prime}{^\top}\!\bs{f^\prime}\}.
\end{equation*}
Exploiting this and \eqref{th2_b} in \eqref{th2_a} leads to the following Lyapunov drift inequality which is evaluated under HD policy given current queue backlogs at slot $n$: 
\begin{gather*}
\mathbb E\bigl\{\Delta W\,\vert\bs q_\circ\bigr\} \leqslant 
\eta\,\mathbb E\{ (\bs a_\circ \!-\! \bs B_\circ\bs{f^\prime} )^{\!\top} \}
\,\mathbb E \{ \bs M_{\!\circ} \} \,\bs q_\circ
+ \mathbb E\bigl\{\Gamma\,\vert\bs q_\circ\bigr\}
\\ 
\Gamma := 
(\bs a_\circ \!-\! \bs B_\circ\bs f^\star )^{\!\top} \bs M_{\!\circ} (\bs a_\circ \!-\! \bs B_\circ\bs f^\star )
+ \frac{\eta}{2}\,\bs{f^\prime}^\top\!\bs{f^\prime}. 
\end{gather*}

Investigating $\Gamma(n)$, note that 
(i) all arrivals are of finite mean and variance, 
(ii) each link actual-transmission is at most equal to the link capacity which is finite, and so both $\bs f^\star(n)$ and $\bs{f^\prime}(n)$ have finite upper bounds, and 
(iii) $\bs M_{\!\circ}(n)$ is a pseudo positive definite matrix in the sense of Lem.~\ref{Lemma1} with finite entries (recall $\phi_{ij}(n)\leqslant1$).
Thus, the expected value of $\Gamma(n)$ is finite at each slot $n$, and so there exists a finite positive scalar $\Gamma_{\!\max}$ such that $\mathbb E\{\Gamma(n)\,\vert\bs q_\circ(n)\} \leqslant \Gamma_{\!\max}$.
Utilizing this in the Lyapunov drift inequality yields
\begin{equation}\label{th2_e}
\mathbb E\bigl\{\Delta W\,\vert\bs q_\circ\bigr\} \leqslant 
\eta\,\mathbb E\{ (\bs a_\circ \!-\! \bs B_\circ\bs{f^\prime} )^{\!\top} \}
\,\mathbb E \{ \bs M_{\!\circ} \} \,\bs q_\circ + \Gamma_{\!\max}\,. 
\end{equation}
In the wake of \eqref{th2_e}, the next theorem is proven by showing that $\mathbb E\{\Delta W\,\vert\bs q_\circ\}$ is always negative for all $\beta\in[0,1]$. 
(Refer to the Appendix for the end of the proof.)

\begin{theorem}[HD Throughput Optimality] 
\label{maximum_throughput}
Over any uniclass wireless network, HD policy with any $\beta\in[0,1]$ is throughput optimal, meaning that it guarantees network stability under all stabilizable arrival vectors.
\end{theorem}

\section{HD Minimum Delay at $\beta=0$}
\label{s:C3_queue_minimum}

Pareto optimal performance of HD policy stands on two pillars: minimization of the average queue congestion $\overline Q$ with $\beta=0$, and minimization of the average routing cost $\overline R$ with $\beta=1$.
This section settles the first pillar based on a timeslot analysis.
The result of this section is analytically proven under the general K\mbox{-}hop interference model, where two wireless links can be activated at the same time if they are at least K$+1$ hops away from one another.
For example, in the 1\mbox{-}hop interference model, links with the exclusive nodes may be scheduled at the same time. 
Let us start with two lemmas (proof in the appendix) that help us analyze the final delay minimization in Th.~\ref{minimum_queue}.

\begin{lemma}\label{Lemma4}
At $\beta=0$ and under the K\mbox{-}hop interference model, timeslot maximization of the functional $D(\bs f,\bs q_\circ,n)$ in \eqref{Dfunction2} is equivalent to timeslot maximization of 
\begin{equation}\label{Dfunction3}
G(\bs f,\bs q_\circ,n) := 2\,\bs f(n)^{\!\top}\bs B\otop\bs q_\circ(n) - \bs f(n)^{\!\top}\bs B\otop\bs B_\circ\bs f(n). 
\end{equation}
\end{lemma}

It is critical to understand that Lem.~\ref{Lemma4} does not claim about the same maximum values for functionals $D$ and $G$, which is obviously not true, but about the same maximizing control action $\bs f(n)$ at each slot $n$. 
Another point is that while at each timeslot, HD maximizes $D$ for all $\beta\in[0,1]$, it maximizes $G$ for only $\beta = 0$.

\begin{lemma}\label{Lemma5}
Consider a uniclass wireless network under an arrival rate $\overline{\bs a_\circ\!}\,$ that is stabilized by a routing policy, resulting in average queue occupancies $\overline{\bs q_\circ\!}\,$ and average link actual transmissions $\overline{\bs f}\,$.
Then the following identity holds:
\begin{equation}\label{Lem4}
\begin{aligned}
2\,\,\overline{\!\cov\{\bs B_\circ\bs f,\bs q_\circ\}\!}\, & 
- \overline{\var\{\bs B_\circ\bs f\}\!}\, =
\\&
2\,\,\overline{\!\cov\{\bs a_\circ,\bs q_\circ \!-\! \bs B_\circ\bs f\}\!}\,+
\overline{\var\{\bs a_\circ\}\!}
\end{aligned}
\end{equation}
where for two random variables $\bs X$ and $\bs Y$, 
$\cov\{\bs X,\bs Y\}:=\mathbb E\{\bs X^{\!\top\!}\bs Y\}-\mathbb E\{\bs X\}^{\!\top}\mathbb E\{\bs Y\}$ and 
$\var\{\bs X\}:=\cov\{\bs X,\bs X\}$.
\end{lemma}

To gain an insight into this lemma, consider a constant arrival vector which makes the right-hand side of \eqref{Lem4} vanished.  
In light of $\cov\{\bs B_\circ\bs f,\bs q_\circ\} = \cov\{\bs f,\bs B\otop\bs q_\circ\}$, equality \eqref{Lem4} then implies that a stabilizing routing decision with a higher average total variance of link forwardings necessarily results in a higher average total covariance between link forwardings and link queue-differentials.
For example, compared with BP that saturates activated links to their capacity limits, HD with a more conservative packet forwarding results in less variations in link actual transmissions.
The lemma then claims that HD leads to a smaller correlation between link forwardings and link queue-differentials, which is confirmed by comparing HD and BP algorithms (see H4 in Sec.~\ref{s:C3_Highlights}).

\begin{definition}
We specify {\it $\cal D$-class routing policies} as a collection of all dynamic routing policies that make timeslot routing decisions based only on current queue occupancies and channel states and so independent of arrival statistics and channel state probabilities.
\end{definition}

By allowing as many routes as possible, $\cal D$-class routing policies tend to distribute traffic all over the network.
This class includes all opportunistic max-weight schedulers that do not incorporate the Markov structure of topology process into their decisions, including BP~\cite{Tassiulas92} and most of its derivations~\cite{Shah06,Dai08,Ross09,Naghshvar12,Baras12,Dai05,Bui11,Shroff11,Huang12,Alresaini12,Moeller10,Martinez11,Srikant09a,Ekici12,Bui09,Jiang11,Srikant09b}. 
The class also encompasses all offline stationary randomized algorithms (possibly unfeasible) that make routing decisions as pure functions only of observed channel states, and so independent of queue occupancies, by typically using the knowledge of arrival statistics and channel state probabilities. 

\begin{theorem}[HD Minimum Delay] 
\label{minimum_queue}
Consider a uniclass wireless network that meets Assum.~\ref{ass2} under a stabilizable arrival rate.
Within $\cal D$-class routing policies and under the K\mbox{-}hop interference model, HD with $\beta=0$ minimizes the average total queue congestion $\overline Q$ as defined in \eqref{average3}, which is proportional to average network delay by Little's Theorem.
\end{theorem}

\section{Classical vs Combinatorial Heat Process}
\label{s:C3_heat}

To formulate heat diffusion on graph, we use the theory of {\it combinatorial geometry}, where the notion of {\it chains-cochains} on a combinatorial domain provides a genuine counterpart for {\it differential forms} in classical geometry.
Details are found in~\cite{Arnold10} and references therein.

\subsection{Heat Equations on Manifolds}

On a smooth manifold $\cal M$ charted in local coordinates~$\bs z$, consider $Q(\bs z,t)$ as spatial distribution of temperature, $\bs F(\bs z,t)$ as heat flux, 
and $A(\bs z,t)$ as scalar field of heat sources (with minus for sinks).
The law of heat conservation entails  
\begin{equation}\label{heat1}
\frac{\partial Q(\bs z,t)}{\partial t} = -\div \bs F(\bs z,t) + A(\bs z,t)\,.
\end{equation}
Fick's law relates the diffusive flux to the concentration, postulating that the heat flux goes from warm regions of high concentration to cold regions of low concentration, with a magnitude that is proportional to the concentration gradient:
\begin{equation}\label{heat2}
\bs F(\bs z,t) = -\sigma(\bs z)\,\nabla Q(\bs z,t)
\end{equation}
where $\sigma(\bs z)$ is {\it thermal diffusivity} that quantifies how fast heat moves through the material.
Combining \eqref{heat1} and \eqref{heat2} together, we obtain  
\begin{equation}\label{heat3}
\frac{\partial Q(\bs z,t)}{\partial t} = \div\bigl(\sigma(\bs z)\,\nabla Q(\bs z,t)\bigr) + A(\bs z,t)\,.
\end{equation}
To have a unique solution, besides time initial condition, one must prescribe $Q$ conditions on a {\it boundary} $\partial{\cal M}$.

\subsection{Heat Equations on Undirected Graphs}
\label{s:C3_heatUndirectedGraphs}

In the context of combinatorial geometry, let us view a graph as a {\it simplicial 1-complex} and transfer elements of classical heat equations to this cell complex. 
In doing so, the smooth manifold~$\cal M$ is replaced by a 0\mbox{-}{\it chain} vector representing the discrete domain, 
the pointwise functions $Q(\bs z,t)$ and $A(\bs z,t)$ are respectively replaced by 0\mbox{-}{\it cochain} vectors $\bs q(t)$ and $\bs a(t)$ (node variables), 
the line integral $\bs F(\bs z,t)$ is replaced by 1\mbox{-}{\it cochain} vector $\bs f(t)$ (edge variable), and 
the thermal diffusivity~$\sigma$ is replaced by a vector of edge weights $\bs\sigma$.

As a 1-complex, the graph structure is fully described by its {\it node-edge} incidence matrix $\bs B$.
\footnote{
The incidence matrix defined in Sec.~\ref{s:C3_key} for a directed graph has the same structure except that the edge directions are substituted for the arbitrarily assigned algebraic topological edge orientations here.} 
Then on an {\it undirected} graph with a node~$d$ as the heat sink, combinatorial analogue of the classical heat equations~\eqref{heat1}--\eqref{heat3} are obtained as 
\begin{gather}
\bs{\dot q}(t) = -\bs B\,\bs f(t) + \bs a(t)
\, , \:\: q_d(t) = 0 
\label{discrete1}
\\
\bs f(t) = \diag(\bs \sigma)\, \bs B^\top \bs q(t)
\label{discrete2}
\\
\bs{\dot q}(t) = 
-\bs B\,\diag(\bs\sigma)\,\bs B^\top \bs q(t) + \bs a(t)
\, , \:\: q_d(t) = 0\,. 
\label{discrete3}
\end{gather}
Note that the boundary~$\partial{\cal M}$ on the manifold collapses to the single node~$d$ on the graph at the fixed zero temperature, which absorbs all the heat generated by the heat sources $\bs a(t)$. 

Enforcing boundary condition $q_d(t)\! = \!0$, one can eliminate the sink~$d$ from \eqref{discrete1}--\eqref{discrete3}, which yields the reduced set of {\it continuous-time} graph heat equations as 
\begin{gather}
\bs f(t) = \diag(\bs \sigma)\, \bs B\otop \bs q_\circ(t)
\label{discrete4}
\\
\bs{\dot q}_\circ(t) = -\bs L_\circ\,\bs q_\circ(t) + \bs a_\circ(t) 
\;,\;
\bs L_\circ := \bs B_\circ\,\diag(\bs\sigma)\, \bs B\otop.
\label{discrete5}
\end{gather}
where as before, subscript $\circ$ denotes a reduced vector or matrix that discards the entries corresponding to the destination node $d$. 
The linear operator $\bs L_\circ$ is called the {\it Dirichlet Laplacian} with respect to the node $d$, which is a symmetric and diagonally dominant matrix.
Further, it can be shown that for any connected graph, $\bs L_\circ$ is {\it positive definite.}

\subsection{Heat Equations on Directed Graphs}

On a directed graph, the combinatorial heat conservation~\eqref{discrete1} remains unchanged, but the Fick's law~\eqref{discrete2} must be modified to allow flow in only one direction.
Let arbitrarily assigned edge orientations concur with edge directions.
Like the undirected case, one can drop the sink node $d$ from equations by fixing $q_d(t) = 0$ as boundary condition.
Then we get the reduced set of {\it continuous-time} heat equations on an {\it uncapacitated directed} graph as
\begin{gather}
\bs f(t) = \diag(\bs \sigma) \max\bigl\{ \bs 0 , \: \bs B\otop \bs q_\circ(t) \bigr\}
\label{discrete6}
\\
\begin{gathered}
\bs{\dot q}_\circ(t) = -\LL_\circ\, \bs q_\circ(t) + \bs a_\circ(t) 
\\
\LL_\circ := \bs B_\circ\,\diag(\bs\sigma)\,\diag\bigl(\II_{\bs B\otop\bs q_\circ(t)\succ\bs 0}\bigr)\bs B\otop. 
\end{gathered}
\label{discrete7}
\end{gather}
We refer to $\LL_\circ$ as {\it nonlinear Dirichlet Laplacian} that acts on a directed graph and, unlike $\bs L_\circ$, is an operand-dependent operator that retains neither linearity nor symmetry.

\begin{remark}
For the first time, heat diffusion on directed graphs is formulated via a nonlinear Laplacian.
This is in agreement with the recent work in~\cite{Ohta09} showing that heat diffusion on Finsler manifolds, the natural counterparts of directed graphs in continuous domain, leads to a {\it nonlinear} Laplacian.
In the graph literature, different {\it linear} Laplacians have been proposed for directed graphs (see \cite[Sec.~3]{Boley11} for a review).
While successful to address some purely graphical issues, they are not able to convey the physics of the diffusion process, nor the intrinsic nonlinearity due to the one-way flow restrictions. 
\end{remark}

Given finite heat sources, heat equations on a connected {\it undirected} graph always lead to finite temperatures at the nodes.
However, for \eqref{discrete7} to have a finite solution, each nonzero heat source needs to connect to the sink through at least one {\it directed} path.
If this basic condition does not hold, the network flow problem has indeed no solution in the sense that there is no way to transfer all commodities, which is heat in our case, to the destination.

\begin{definition}\label{def_feasible}
A nonzero heat source is {\it feasible} if it connects to sink through at least one directed path, with the path being directed from source to sink for a positive heat source and from sink to source for a negative heat source.
A vector of heat sources is feasible if each of its nonzero components is feasible.
\end{definition}

\section{Wireless Network Thermodynamics}
\label{s:C3_thermodynamics}

Though defined on a directed graph, the heat equations \eqref{discrete6}--\eqref{discrete7} still represent a deterministic, continuous-time process with no link interference.
The latter, particularly, makes the wireless problem quite intractable.
Nonetheless, this section advocates a genuine diffusion on stochastic, slotted-time, interference networks by showing that under HD routing policy, the long-term average dynamics of an interference wireless network comply with non-interference combinatorial heat equations on a suitably-weighted directed graph.

\subsection{HD Fluid Limit}

Fluid limit of a stochastic process is the limiting dynamics obtained by {\it scaling} in time and amplitude.
Under very mild conditions, it is shown that these scaled trajectories converge to a set of deterministic equations called {\it fluid model.}
Using such a deterministic model, one can analyze {\it rate-level,} rather than {\it packet-level,} behavior of the original stochastic process.
Details are found in~\cite{Dai99,Bramson08} and references therein.

{\it Fluid limit:} 
Let $\bs X(\omega,t)$ be a realization of a continuous-time stochastic process $\bs X$ along a sample path $\omega$.
Define the scaled process $\bs X^r(\omega,t):=\bs X(\omega,rt)/r$ for any $r>0$.
A~deterministic function $\bs{\tilde X}(t)$ is a {\it fluid limit} if there exist a sequence~$r$ and a sample path~$\omega$ such that $\;\lim_{r\to\infty}\bs X^r(\omega,t)\to\bs{\tilde X}(t)$ uniformly on compact sets.
For a stable flow network, the existence of fluid limits is guaranteed if exogenous arrivals are of finite variance.
It is further shown that each fluid limit is Lipschitz-continuous, and so differentiable, almost everywhere with respect to Lebesgue measure on $[0,\infty)$.

{\it Cumulative process:} 
Note that the fluid theorem is defined for continuous-time stochastic processes, while a wireless network is a slotted-time process.
To resolve this issue, we derive a first-order continuous-time approximation of wireless network dynamics using its cumulative model.
Let $\bs a_\circ^\tot(n)$ and $\bs f^\tot(n)$ be respectively the vector of cumulative node arrivals and link transmissions up to slot~$n$.
In light of dynamic equation \eqref{model5} and by assuming the initial conditions $\bs a_\circ^\tot(0)\!=\!\bs 0$ and $\bs f^\tot(0)\!=\!\bs 0$, we obtain
\begin{equation}\label{fluid01}
\bs q_\circ(n)=\bs q_\circ(0) + \bs a_\circ^\tot(n) - \bs B_\circ\bs f^\tot(n).
\end{equation}
Let $\smash{\widehat{f_{ij}}}(n)$ be the predicted number of packets that link~$ij$ would transmit if it were activated at slot~$n$ and compose the vector $\smash{\smash{\widehat{\bs f(}}}n)$.
Also let $T_{\bs\pi}(n)$ be the cumulative number of timeslots, until slot $n$, in which the scheduling vector $\bs\pi\!\in\!\Pi$ has been selected.
Assuming the initial condition $T_{\bs\pi}(0)=0$, one can verify that
\begin{equation}\label{fluid02}
\bs f^\tot(n) =\! \sum_{\bs\pi\in\Pi}\,\sum_{k=1}^{n}
\bigl( T_{\bs\pi}(k)-T_{\bs\pi}(k\!-\!1) \bigr)
\bigl( \bs\pi \odot \smash{\smash{\widehat{\bs f(}}}k) \bigr).
\end{equation}
The first parenthesis equals 1 if the scheduling vector $\bs\pi$ has been selected at slot $k$, and 0 otherwise.
The term $(\bs\pi \odot \smash{\smash{\widehat{\bs f(}}}k))$ represents the number of packets that could be transmitted over each link if the scheduling vector $\bs\pi$ were selected. 
Note that a routing policy needs to determine each entry of $\smash{\widehat{\bs f(}}k)$ and select a scheduling vector $\bs\pi\!\in\!\Pi$ at each timeslot.

{\it General fluid equations:}
Given a sample path $\omega$, we extend a slotted-time process to be continuous-time via linear interpolation in each timeslot interval $(n,n+1)$.
With no loss of generality, let exogenous arrivals occur at the beginning of each timeslot so that $\bs a_\circ^\tot(t)$ represents cumulative arrivals by time~$t$. 
Assuming normalized timeslots with the period of time unit, \eqref{fluid01} directly provides the first set of stochastic general fluid equations as
\begin{gather}
\bs q_\circ(t)=\bs q_\circ(0) + \bs a_\circ^\tot(t) - \bs B_\circ\bs f^\tot(t) 
\label{fluid1}
\\
\bs a_\circ^\tot(t) = \overline{\bs a_\circ\!}\:\,t
\label{fluid4a}
\end{gather}
with $\overline{\bs a_\circ\!}\,$ being the time average expectation of the random arrivals $\bs a_\circ(n)$.
The second set of general fluid equations are obtained from the time derivative of \eqref{fluid02} as
\begin{gather}
\bs{\dot f}{^\tot}(t) =
\sum\nolimits_{\bs\pi\in\Pi}
\dot{T}_{\bs\pi}(t)
\bigl( \bs\pi \odot \smash{\widehat{\bs f(}}t) \bigr) 
\label{fluid2}
\\
{\underset{\bs\pi\in\Pi}{\dot{T}_{\bs\pi}(t)} = 
\begin{cases}
1 	& 		\text{if $\bs\pi$ is chosen at time $t$}
\\
0 &			\text{otherwise}
\end{cases}} 
\label{fluid3a}
\\
\sum\nolimits_{\bs\pi\in\Pi}T_{\bs\pi}(t) \!=\! t \:\text{ with $T_{\bs\pi}(t)$ nondecreasing.}
\label{fluid3}
\end{gather}
%
Note that \eqref{fluid2} entails the existence of a $\delta>0$ such that 
\begin{equation*}
f_{ij}^\tot(t^\prime) - f_{ij}^\tot(t) = 
\sum\nolimits_{\bs\pi\in\Pi} \!
\pi_{ij} \;
\smash{\widehat{f_{ij}}}(t)
\bigl( T_{\bs\pi}(t^\prime) - T_{\bs\pi}(t) \bigr)
\end{equation*}
for any $t^\prime\!\in[t,t+\delta]$.
This states the fact that if a link has a positive flow of packets at time~$t$, the number of packets transmitted by the link in an interval $[t,t^\prime]\subset[t,t+\delta]$ is equal to the amount of time the link has been activated during~$[t,t^\prime]$ multiplied by its transmission rate prediction at time $t$.

{\it Particular fluid equations:}
While \eqref{fluid1}--\eqref{fluid3} hold for any stable network operating under an arbitrary non-idling control policy, each policy determines $\smash{\smash{\widehat{\bs f(}}t)}$ and $T_{\bs\pi}(t)$ in its own particular way.
Referring to~\eqref{HD2}, HD policy enforces  
\begin{equation}
\smash{\widehat{\bs f(}}t) \overset{\mathrm{HD}}{=}
\min\bigl\{ \,\overline{\bs\Phi}\bigl(\bs B\otop \bs q_\circ(t)\bigr){^+},\,\overline{\bs\mu}\, \bigr\}.
\label{fluid4}
\end{equation}
where $\overline{\bs\Phi}$ represents the time average expectation of $\bs\Phi(n)$ as defined in \eqref{phi}.
Note that the existence of $\overline{\bs\mu}$ and $\overline{\bs\rho}$ is secured by \eqref{topology1} and \eqref{topology2}. 
Referring now to~\eqref{HD3} and \eqref{HD4}, HD policy determines the scheduling vector $\bs\pi(t)$ by solving the following max-weight optimization problem: 
\begin{gather}
\bs\pi(t) = \arg\max\nolimits_{\bs\pi\in\Pi} \; 
\bs\pi\!^\top \bs w(t)
\label{fluid52}
\\
\bs w(t) \overset{\mathrm{HD}}{=}\, 
\smash{\widehat{\bs f(}}t) \odot
\bigl( \,2\;\overline{\bs\Phi}\,\bs B\otop \bs q_\circ(t) - \smash{\widehat{\bs f(}}t) \,\bigr) 
\label{fluid51}
\end{gather}
where $\bs w(t)$ is the vector of weights assigned by HD policy to each link at time~$t$.

For a comparison, observe that the original BP solves the same max-weight optimization problem \eqref{fluid52} to find a scheduling vector $\bs\pi(t)$, but it enforces $\smash{\widehat{\bs f(}}t)$ and $\bs w(t)$ to be 
\begin{gather}
{\smash{\widehat{f_{ij}}}(t) \overset{\mathrm{BP}}{=}
\begin{cases}
\min\{ q_i(t), \,\overline{\!\mu_{ij}\!}\, \} & \text{if } q_{ij}(t)>0
\\[3pt]
0 &			\text{otherwise}
\end{cases}}  
\label{fluid53}
\\
\bs w(t) \overset{\mathrm{BP}}{=}\,
\overline{\bs\mu} \odot
\bigl( \bs B\otop \bs q_\circ(t) \bigr){^+}.
\label{fluid54}
\end{gather}

\begin{theorem}[HD Fluid Model]\label{prop_fluid_model1} 
On a uniclass wireless network stabilized by Pareto optimal HD policy, every fluid limit 
$\bs{\tilde X}(t) = \bigl( \bs{\tilde q}_\circ(t),\bs{\tilde f}{^\tot(t)},{\tilde T}_{\bs\pi}(t) \bigr)$ satisfies {\it HD fluid model,} which is defined as the collection of deterministic continuous-time equations~\eqref{fluid1}--\eqref{fluid51}.
\end{theorem}

\begin{remark}
It is important to discriminate between fluid limit and fluid model of a discrete-time stochastic process.
The former is the scaled process of the first-order continuous-time approximation for an arbitrary realization of the stochastic process, while the latter is a (set of) fully deterministic, continuous-time equation(s). 
Consider now a wireless network under HD routing policy, where the discrete-time stochastic processes $\bs q_\circ(n)$, $\bs f{^\tot(n)}$ and $T_{\bs\pi}(n)$ have respectively the continuous-time fluid limits $\bs{\tilde q}_\circ(t)$, $\bs{\tilde f}{^\tot(t)}$ and ${\tilde T}_{\bs\pi}(t)$.
Then Th.~\ref{prop_fluid_model1} states that for large enough scaling factors, the fluid limit of every realization converges to a set of deterministic, continuous-time functions $\bs q_\circ(t)$, $\bs f{^\tot(t)}$ and $T_{\bs\pi}(t)$ which solve the HD fluid model equations~\eqref{fluid1}--\eqref{fluid51}.
\end{remark}

%

\subsection{Thermodynamic-Like Packet Routing}
\label{s:C3_thermodynamic_like}

Consider a uniclass wireless network with packets being routed under HD policy (microscopic flow). 
At each timeslot, HD policy activates a particular set of links to transmit a specific number of packets over them. 
Obviously, each link transmits packets at some slots and is switched off at some other slots. 
Let us now look at the limit flow on each link, defined as the total number of packets transmitted over the link during a large period of time divided by the time duration. 
We claim that observing average packet flow in limit (macroscopic flow), it takes the form of heat flow on the underlying directed graph with suitably-weighted edges.

Consider a thermal graph with the same node-edge incidence matrix $\bs B_\circ$ and the edge thermal diffusivity $\sigma_{ij}=\,\overline{\!\phi_{ij}\!}\,$. 
Associate with each arrival $a_i(n)$ on the wireless network a static heat source of intensity $\overline{a_i}$ on the graph and fix zero temperature at the destination node.
The flow of heat on this directed graph is governed by \eqref{discrete6}\mbox{--}\eqref{discrete7},
which provides the wireless network with a static {\it reference thermal model} as
\begin{gather}
\bs f^\rf = \overline{\bs\Phi} \, \max\bigl\{ \bs 0 , \: \bs B\otop \bs q_\circ^\rf \bigr\}
\label{discrete8}
\\
\overline{\bs a_\circ\!}  = \LL{_\circ^\rf} \bs q_\circ^\rf 
\: , \:
\LL{_\circ^\rf} := \bs B_\circ\overline{\bs\Phi} \, 
\diag\bigl(\II_{\bs B\otop \bs q_{\circ\!}^\rf\succ\bs 0}\bigr)
\bs B\otop.
\label{discrete9}
\end{gather}
Note that $\,\overline{\!\phi_{ij}\!}\,$ depends not only on the link cost factor $\,\overline{\!\rho_{ij}\!}\,$, but also on the penalty factor $\beta$, where varying $\beta$ leads to different edge weights and so different graph topologies.

Recall that $T_{\bs\pi}(t)$ represents the cumulative time until $t$ in which the scheduling vector $\bs\pi\in\Pi$ has been selected.
Obviously, each scheduling policy leads to its own specific $T_{\bs\pi}(t)$.  
For example, under HD policy, $T_{\bs\pi}(t)$ is determined by the HD scheduling \eqref{fluid52}--\eqref{fluid51}, while the original BP determines it according to \eqref{fluid53}--\eqref{fluid54}.

\begin{definition}
Under a sequence of wireless link scheduling, the {\it effective capacity} on each link is the time average expectation of capacity made genuinely available on that link: 
\begin{equation*}\label{effCapacity}
\bs\mu_\eff :=
\limsup_{\tau\to\infty} \frac{1}{\tau} 
\sum_{\bs\pi\in\Pi} \sum_{n=0}^\tau 
\bigl( T_{\bs\pi}(n)-T_{\bs\pi}(n-1) \bigr)
\bigl( \bs\pi \odot \mathbb E\{\bs\mu(n)\} \bigr)
\end{equation*}
where $\bs\mu_\eff$ denotes the vector of effective link capacities.
\end{definition}

Observe that the classical heat equations \eqref{heat1}--\eqref{heat3}, and their combinatorial counterparts \eqref{discrete4}--\eqref{discrete5}, take no limit in either flow direction or flow capacity.
Then note that while \eqref{discrete6}--\eqref{discrete7} extend heat equations to directed graphs, they still consider no capacity limits on branches. 
In fact, the underneath assumption is that the flow of heat on each directed edge follows the Fick's law of diffusion, not intervened by the edge capacity.
 
\begin{assumption}\label{ass1}
Given an arrival rate $\overline{\bs a_\circ\!}\,$, there exists at least one sequence of wireless link scheduling under which the effective link capacities meet the requirement of reference heat flow \eqref{discrete8}, which is stated by $\bs f^\rf \! \preccurlyeq \bs\mu_\eff$.
\end{assumption}   

While $\bs\mu_\eff$ is a network characteristic and independent of arrivals, satisfaction of Assum.~\ref{ass1} does depend on arrivals. 
Further, for a given arrival rate, there could be a large number of link scheduling sequences that meet the requirement.

\begin{theorem}[Wireless Network Thermodynamics]\label{prop_fluid_model2}
Consider a uniclass wireless network that meets Assum.~\ref{ass2} and \ref{ass1} under a stabilizable arrival rate.
Then the HD fluid model \eqref{fluid1}--\eqref{fluid51} asymptotically converges to the thermal model \eqref{discrete8}--\eqref{discrete9}. 
In particular, HD fluid model with $\beta=0$ complies with heat equations on an unweighted directed graph, and with $\beta=1$ to those on a weighted directed graph with $\sigma_{ij}=1/\,\overline{\!\rho_{ij}\!}\,$. 
\end{theorem}

\begin{remark}\label{remark6}
Assum.~\ref{ass1} examines if it is possible in principle to stabilize the wireless network such that its fluid limit follows {\it uncapacitated} heat equations.
We fully revoke this assumption in \cite{RezaACM} by developing diffusion equations on {\it capacitated} directed graphs and showing that the fluid equations \eqref{fluid1}--\eqref{fluid51} still respect them with no need of satisfying Assum.~\ref{ass1}.
In fact, we solve a more complicated diffusion problem in \cite{RezaACM}, where not only directed edges have limited capacity, but flows with different destinations need to be carried over the network, which raises the challenge of optimal designation of edge capacities to each of them.
\end{remark}

\section{HD Minimum Routing Cost at $\beta=1$}
\label{s:C3_cost_minimization}

To establish the second pillar of HD Pareto optimality, this section shows, via Dirichlet's principle, that average quadratic routing cost is minimized under HD policy with $\beta\!=\!1$.
In fact, we show a more general result that HD with any $\beta\in[0,1]$ solves the following $\beta$-dependent optimization problem:
\begin{equation}\label{average21}
\mathrm{Minimize}\;\;\;\, 
\sum\nolimits_{ij\in\cal E}\;\overline{\!(f_{ij})^2/\phi_{ij} }
\end{equation}
where $\beta\!=\!1$ leads to $\phi_{ij}=1/\rho_{ij}$, which recovers \eqref{average2} on minimizing the average quadratic routing cost $\overline{R}$. 

\begin{remark}
At $\beta\!=\!0$, we get $\phi_{ij}=0.5$ for all links, implying that 
$\sum\nolimits_{ij}\overline{\!(f_{ij})^2}$ is minimized by HD with $\beta\!=\!1$.
The average total queue congestion $\overline Q$ is also minimized by HD with $\beta=0$ (see Th.~\ref{minimum_queue}). 
This entails that the two objective functions are minimized by the same timeslot control action $\bs f(n)$, which makes the ground for our results on HD weak Pareto optimality for uniform link costs in Sec.~\ref{s:C3_pareto}. 
\end{remark}

\subsection{Classical Dirichlet Principle}

Consider the classical heat diffusion equations~\eqref{heat1}--\eqref{heat3} subject to constant heat sources $A(\bs z)$. 
In a steady-state thermal conduction, the amount of heat entering any region of manifold $\cal M$ is equal to the amount of heat leaving out the region. 
Thus, while partial derivatives of temperature with respect to space may have either zero or nonzero values, all time derivatives of temperature at any point on $\cal M$ will remain uniformly zero. 
This leads to the classical Poisson equation 
\begin{equation*}
\div\bigl(\sigma(\bs z)\,\nabla Q(\bs z)\bigr) + A(\bs z) = 0
\end{equation*}
which formulates stationary heat transfer by substituting zero for the time derivative of temperature in~\eqref{heat3}.
Dirichlet's principle then states that the Poisson equation has a unique solution that minimizes the Dirichlet energy
\begin{equation*}
E_D\bigl(Q(\bs z)\bigr) := \int_{\cal M} \Bigl(\,\frac{1}{2}\,\sigma(\bs z)\lVert\nabla Q(\bs z) \rVert^2 - Q(\bs z) A(\bs z) \Bigr){\mathrm d}\bs z
\end{equation*}
among all twice differentiable functions $Q(\bs z)$ that respect the boundary conditions on $\partial{\cal M}$.

\subsection{Combinatorial Dirichlet Principle}

To derive the combinatorial analogue of Poisson equation on {\it undirected} graphs, one identifies the classical $\div$ with the boundary operator $\bs B$ and the classical gradient $\nabla$ with the minus\footnote{
In vector calculus, the gradient of a scalar field is positive in the direction of increase of the field.
On a graph, on the other hand, we take the gradient of a node variable positive in the direction of decrease of the variable. 
By the same reason, the classical Laplace operator is a negative semi-definite operator, while the graph Laplacian is a positive semi-definite matrix.
}
of coboundary operator $\bs B^\top\!$.
Fixing $q_d(t) = 0$ yields 
\begin{equation}\label{Poisson1}
-\bs L_\circ\,\bs q_\circ + \bs a_\circ = \bs 0
\end{equation}
which correctly realizes \eqref{discrete5} in steady-state. 
Note that the equation has no time variable $(t)$, since it represents the steady-state condition. 
It is not difficult to see that, like the classical case, this equation has a unique solution that minimizes the combinatorial Dirichlet energy
\begin{equation*}\label{Poisson2}
E_D(\bs q_\circ) := 
\frac{1}{2}\; \bs q\otop \bs L_\circ \, \bs q_\circ - \bs q\otop \bs a_\circ\,.
\end{equation*}
The proof of Dirichlet's principle is much simpler in the combinatorial case. 
In fact, as $\bs L_\circ$ is positive definite, $E_D(\bs q_\circ)$ is convex and so has a minimum at the critical point where its first order variation vanishes, which readily leads to the combinatorial Poisson equation~\eqref{Poisson1}.

\subsection{Nonlinear Dirichlet Principle}

Essentially, the Poisson equation on a {\it directed} graph should capture the steady-state behavior of combinatorial nonlinear diffusion process~\eqref{discrete7} subject to constant heat sources $\bs a_\circ$.
This leads to the following nonlinear Poisson equation: 
\begin{equation}\label{Poisson3}
-\LL_\circ\,\bs q_\circ + \bs a_\circ = \bs 0\,.
\end{equation}
Difficulty arises from the fact that contrary to linear Laplacian $\bs L_\circ$ on undirected graphs that is a symmetric positive definite matrix, $\LL_\circ$ is an operand-dependent operator that retains neither linearity nor symmetricity.
Thus, the easy way of proving Dirichlet's principle on undirected graphs ceases to exist here,
as one can not claim that $\LL_\circ\bs q_\circ$ in~\eqref{Poisson3} is the directional derivative of $\frac{1}{2}\,\bs q\otop\LL_\circ\bs q_\circ$.
Nonetheless, we extend the concept of combinatorial Dirichlet principle to directed graphs by the next theorem.

\begin{theorem}[Nonlinear Dirichlet Principle]\label{prop_fluid_model3}
Given a feasible $\bs a_\circ$ on a directed graph, the nonlinear Poisson equation~\eqref{Poisson3} has a unique solution that minimizes the nonlinear Dirichlet energy
\begin{equation}\label{Poisson4}
\vec E_D(\bs q_\circ) := 
\frac{1}{2}\; \bs q\otop \LL_\circ\, \bs q_\circ - \bs q\otop \bs a_\circ\,.
\end{equation}
\end{theorem}

\begin{remark}
Though Dirichlet's principle on undirected graphs has been known for long time, its extension to directed graphs is completely new to literature. 
As a model of heat flow on directed graphs, one can conceptualize a resistive network with a diode added to each edge \cite{RezaACM}. 
Electrical current -- the counterpart of combinatorial heat flux -- moves along negative gradient of voltage, but only under the condition of respecting the diode direction. 
Another example is a piping network of liquid/gas with a check valve on each line.
Again, the liquid/gas flows along negative gradient of pressure, while each check valve allows the flow in only one direction. 
\end{remark}

\subsection{Quadratic Routing Cost Minimization}

The framework of Th.~\ref{prop_fluid_model3} is not yet aligned with what we need for the optimization problem~\eqref{average21}.
The next theorem resolves this incongruity by showing that minimizing the functional~\eqref{Poisson4} is indeed the dual of minimizing network energy dissipation, known as Thomson's principle, on the directed graph with zero duality gap.

\begin{theorem}[Nonlinear Thomson Principle]
\label{prop_fluid_model5}
Minimizing the nonlinear Dirichlet energy~\eqref{Poisson4} subject to the nonlinear Poisson equation~\eqref{Poisson3} is equivalent to minimizing total energy dissipation on the graph subject to flow conservation at the nodes, stated by
\begin{equation}\label{Poisson7}
\begin{aligned}
\min\nolimits&_{\bs f\succcurlyeq \bs 0} & &\vec E_R(\bs f) := 
\bs f^\top \diag(\bs\sigma)^{-1} \bs f
\\
\mathrm{s.t.}& & & \bs B_\circ\bs f = \bs a_\circ\,
\end{aligned}
\end{equation}
where $\bs f\!\succcurlyeq \bs 0$ is imposed by network directionality.
Further, temperatures at the nodes play the natural role of the Lagrange multipliers in the dual of the optimization problem \eqref{Poisson7}.
\end{theorem}

It is worth comparing the minimization problem \eqref{Poisson7} with the celebrated law of least energy dissipation on resistive networks. 
In essence, Th.~\ref{prop_fluid_model5} extends the law to directed graphs, or to nonlinear resistive-diode networks for that matter \cite{RezaACM}.
The upshot is then due to the connection between heat diffusion on capacitated directed graphs and HD fluid limit, which brings together circuit theory and wireless networking under one umbrella. 

\begin{theorem}[HD Minimum Routing Cost]\label{prop_fluid_model4}
Consider a uniclass wireless network that meets Assum.~\ref{ass2} and \ref{ass1} under a stabilizable arrival rate.
Then HD policy solves the $\beta$-dependent optimization problem~\eqref{average21}. 
In particular, HD policy with $\beta\!=\!1$ minimizes the quadratic routing cost $\overline R$ as defined in \eqref{average2}.
\end{theorem}

In light of Th.~\ref{prop_fluid_model2}, every expected time average value on a stochastic wireless network governed by HD policy follows the corresponding stationary value produced by nonlinear heat equations on the suitably weighted underlying directed graph.
In particular, the $\beta$-dependent objective function in~\eqref{average21} complies with the total energy dissipation $\vec E_R(\bs f)$ on the graph weighted by $\sigma_{ij}=\,\overline{\!\phi_{ij}\!}\;$.
By the same token, the average quadratic routing cost $\overline R$ complies with $\vec E_R(\bs f)$ on the graph weighted by $\sigma_{ij}=1/\,\overline{\!\rho_{ij}\!}\;$.

\begin{remark}
As Rem.~\ref{remark6} explained, Assum.~\ref{ass1} ensures that the link capacity constraints of wireless network do not intervene the Fick's law on its underlying directed graph. 
Again, this assumption is fully revoked in \cite{RezaACM, RezaThermo} by developing Dirichlet's principle on {\it capacitated} directed graphs and showing that HD fluid model still complies with it.
\end{remark}

\section{Pareto Optimality}
\label{s:C3_pareto}

Minimizing average network delay and minimizing average routing cost are often conflicting objectives, meaning that as one decreases the other has to increase. 
This naturally leads to a multi-objective optimization framework.
Then the favorite operating points lie on the {\it Pareto boundary} that corresponds to equilibria from which any deviation results in performance degradation in at least one objective. 
In other words, a Pareto optimal solution is a state of allocation of resources from which it is impossible to reallocate so as to make any one objective better off without making at least another objective worse off.

\subsection{Strong Pareto Optimality for Nonuniform Link Costs}

We have shown that HD with $\beta=0$ minimizes the average network delay $\overline Q$ among all $\cal D$-class routing policies -- solving the optimization problem \eqref{average3}.
We have also shown that HD with ${\beta=1}$ strictly minimizes the quadratic routing cost $\overline R$ among all stabilizing routing algorithms -- solving the optimization problem \eqref{average2}.
Consider now the region of operation built on joint variables $(\overline{Q},\overline{R})$ in which $\overline Q$ is achievable by $\cal D$-class routing policies (possibly unfeasible). 
The next theorem shows that HD policy operates on the Pareto boundary of this $(\overline{Q},\overline{R})$ region by altering $\beta\in[0,1]$ -- solving the multi-objective optimization problem~\eqref{average4}. 

\begin{theorem}[HD Pareto Optimality]\label{Pareto_performance}
Consider a uniclass wireless network that meets Assum.~\ref{ass2} and \ref{ass1} under a stabilizable arrival rate and the K\mbox{-}hop interference model.
Suppose that the operating region built on all possible joint variables~$(\overline Q\,,\overline R)$ with $\overline Q$ produced by a $\cal D$-class routing policy is convex. 
Then HD policy operates on the Pareto boundary of $(\overline Q\,,\overline R)$ region by altering $\beta\in[0,1]$.
\end{theorem}

It is worth to note that in the case of non-convex Pareto boundary, HD with $\beta\in[0,1]$ still covers the points on convex parts of the boundary, though some Pareto optimal points lie on non-convex parts~\cite{Stadler84, Koski88, Kim05}.
The convexity is jeopardized in the presence of a positive correlation between $\overline Q$ and $\overline R$, e.g., if the routing cost is defined as it could grow by the increase of  queue occupancy. 

\begin{remark}
To the best of our knowledge, this is the first time a {\it network layer} routing policy provides Pareto optimal performance with respect to average network delay and routing cost, without requiring any knowledge of traffic and topology.
\end{remark}

\begin{figure}[t!]
	\centering
	\includegraphics[ scale=0.8, trim=5.5cm 5.4cm 10.5cm 7.5cm, clip=true ]{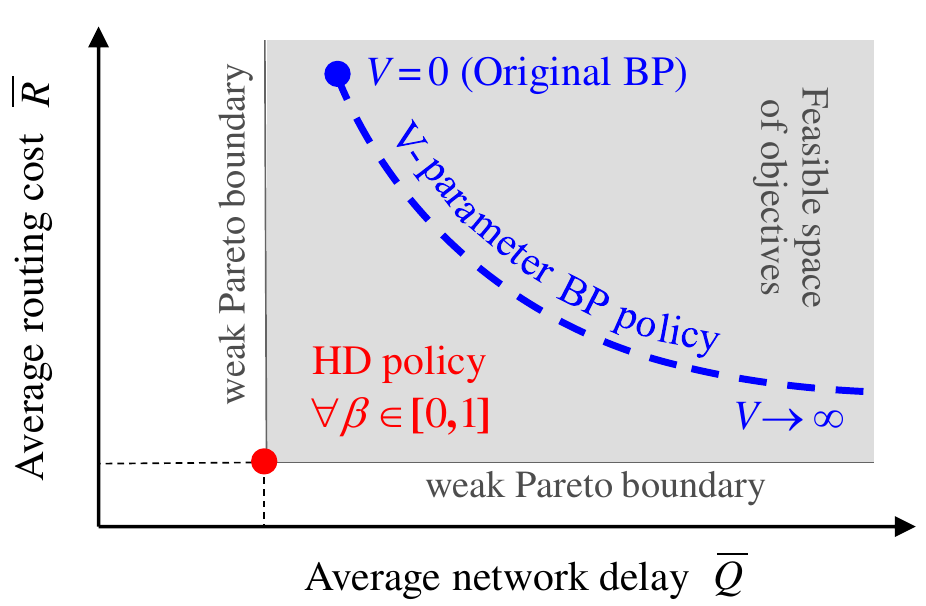}  
	\caption{
	Graphical description of weak Pareto boundary with respect to average queue congestion and the Dirichlet routing cost when cost factor for all links connected to the final destination converge to one and for all other links converge to two, contrasting the performance of HD with V-parameter BP.
	}
\label{figureWeak}
\end{figure}

\subsection{Weak Pareto Optimality for Uniform Link Costs}

Recall from \eqref{HD2} that $\phi_{ij}(n) = (1\! - \!\beta)/\vartheta_{ij} + \beta/\rho_{ij}(n)$ with $\vartheta_{ij} = 1$ if node $j$ is the final destination, i.e., $j=d$, and $\vartheta_{ij} = 2$ otherwise.
When the cost factors in all links converge to $\vartheta_{ij}$, we get $\phi_{ij}=1/\vartheta_{ij}$ for any $\beta$, and so the performance of HD policy turns to be independent of the penalty factor~$\beta$. 
Considering this observation along with Th.~\ref{minimum_queue} implies that the average network delay~$\overline Q$ must be minimized for all~$\beta\in[0,1]$.
Considering it along with Th.~\ref{prop_fluid_model4}, on the other hand, implies that the routing cost~$\overline R$ must also be minimized for all~$\beta\in[0,1]$.
Holding these two requirements at the same time entails that $\overline Q$ and $\overline R$ must be minimized together, which equivalently means that the Pareto boundary of $(\overline Q\,,\overline R)$ region must shrink into one single point. 
Such an operating point is called {\it weakly} Pareto optimal in the sense that no tradeoff is allowed as it is impossible to strictly improve at least one operating objective.
The upshot is formalized by the next corollary.

\begin{corollary} 
Consider a uniclass wireless network under the same condition of Th.~\ref{Pareto_performance}. 
Suppose the cost factors for all wireless links converge to $\vartheta_{ij}$, defined in \eqref{HD2}.
Then the Pareto boundary of $(\overline Q\,,\overline R)$ region shrinks to a point at which HD policy operates for all~$\beta\in[0,1]$.
\end{corollary}

Under uniform cost factor condition for all links, Fig.~\ref{figureWeak} provides a graphical illustration of the feasible region built on 
$(\overline Q\,,\overline R)$.
It emphasizes HD operation at the weakly Pareto optimal point for all $\beta\in[0,1]$ in comparison with the performance of V-parameter BP for $V\in[0,\infty)$.

\section{Conclusion}
\label{s:C3_conclusion}

We have introduced a network layer routing policy, called Heat-Diffusion~(HD), for uniclass wireless networks that 
(i)~is throughput optimal,
(ii)~minimizes average quadratic routing cost,
(iii)~minimizes average network delay within an important class of routing policies,
(iv)~provides a Pareto optimal tradeoff between average network delay and quadratic routing cost, and
(v)~enjoys the same algorithmic structure, complexity and overhead as Back-Pressure~(BP) routing policy.
Further, HD policy is strongly connected to the world of heat calculus in mathematics, which we believe opens the door to a rich array of theoretical techniques to analyze and optimize wireless networking. 
For example, such a connection provides a new way of analyzing the impact of wireless network topology on stability and capacity region~\cite{Chi14} or on delay/routing energy performance \cite{Chi16}.
A decentralized HD protocol has been pragmatically implemented and experimentally evaluated in \cite{HDCP16} for data collection in wireless sensor networks, including a comparative analysis of its performance with respect to the Backpressure Collection Protocol \cite{Moeller10}.
In \cite{JuanReza}, a HD-based delay-aware framework is designed for joint dynamic routing and link-scheduling in multihop wireless networks.

Though motivated by wireless networks, the HD framework can be extended in various ways to other application areas.
Among them is packet scheduling in high speed switches with a lot of attention in recent years. 
Resource allocation problems in manufacturing and transportation also fall within the scope of the model we considered here.

\section*{Acknowledgment}
The authors thank Professor Michael Neely at USC, for his helpful feedback and discussions on this work. 

\section*{Appendix A\\Proof of Theorems and Lemmas}
\label{s:appendixA}

Note that in the proofs we often drop timeslot variable $(n)$ for ease of notation and concision.

\subsection*{Proof of Theorem~\ref{key_property} (HD Key Property)}

One can verify that
\begin{equation*}
D(\bs f,\bs q_\circ,n) = \sum\nolimits_{ij\in\cal E} 2\,\phi_{ij}(n)q_{ij}(n)f_{ij}(n) - f_{ij}(n)^2.
\end{equation*}
Let us temporarily relax all network constraints. 
Then each link-related component of $D(\bs f,\bs q_\circ,n)$ turns to be strictly concave. 
For each link $ij$, by taking the first derivative with respect to $f_{ij}$, we find the maximizing link transmission $f_{ij}^\opt\! = \phi_{ij}q_{ij}$.
Considering the link constraints that $f_{ij}$ must be non-negative and at most equal to the link capacity yields
\[
f_{ij}^\opt\! = \min\{ \phi_{ij}\,q_{ij}{^+} ,\: \mu_{ij} \}
\]
which follows $\smash{\widehat{f_{ij}}}$ in~\eqref{HD2}.
Considering the link interference constraint, on the other hand, enforces to activate the links that contribute most to the $D$ maximization.
Then assuming that an interference model does not let a node transmit to more than one neighbor at the same time, the latter directly leads to the max-weight scheduling~\eqref{HD4} alongside the HD weighting~\eqref{HD3}, which concludes the proof.
\QEDA

\subsection*{Proof of Lemma~\ref{Lemma1}}

Define $\bs\Delta := \bs B_\circ\bs B\otop$ and $\bs\Delta_\phi := \bs B_\circ\bs\Phi\bs B\otop$, which are both positive definite matrices.
Since $\bs\Delta_\phi^{\!1/2}\bs\Delta^{\!-1}\bs\Delta_\phi^{\!1/2}$ is congruent to $\bs\Delta^{\!-1}$ which has only positive eigenvalues, by Sylvester's law of inertia, $\bs\Delta_\phi^{\!1/2}\bs\Delta^{\!-1}\bs\Delta_\phi^{\!1/2}$ has only positive eigenvalues too.
The latter is similar to $\bs M_{\!\circ}$, and so they have the same eigenvalues, proving that $\bs M_{\!\circ}$ has only positive eigenvalues.

We now show that $\bs x^{\!\top}\bs M_{\!\circ}\,\bs x \geqslant 0$.
Letting $\bs v := \bs\Delta^{\!-1} \bs x$ and substituting for $\bs M_{\!\circ}$, it suffices to show that
\begin{equation}\label{ProofLem1_1}
(\bs B\otop\bs v)\!^\top\bigl(\bs\Phi\bs B\otop\bs B_\circ\bigr)(\bs B\otop\bs v) \geqslant 0\,.
\end{equation}
Doing another change of variable, let $\bs f := \bs B\otop\bs v$ that represents an edge vector in which $f_{ij} = v_i - v_j$, $\forall\, ij\in\cal E$.
Recall that $\bs B_\circ$ is a signed node-edge incidence matrix with arbitrarily assigned algebraic topological edge orientations. 
Let us assign edge orientations such that $f_{ij}\! \geqslant 0$, $\forall\, ij\in\cal E$.
Then to fulfill \eqref{ProofLem1_1}, it suffices to show that $\bs f^\top \bs\Phi\bs B\otop\bs B_\circ \, \bs f \geqslant 0$ subject to $\bs f \succcurlyeq \bs 0$, which reads $f_{ij}\! \geqslant 0$, $\forall\, ij\in\cal E$.
To this end, we equivalently show that minimum cost in the following optimization problem is non-negative: 
\begin{equation*}\label{ProofLem1_2}
\min_{\bs f \succcurlyeq \bs 0}\;\;\bs f^\top \bs\Phi\bs B\otop\bs B_\circ \, \bs f\,.
\end{equation*}
Let us construct the Lagrangian dual problem 
\begin{equation}\label{ProofLem1_3}
\max_{\bs\lambda\succcurlyeq\bs 0}\;\min_{\bs f}\;
\Bigl(\, {\cal L}(\bs\lambda,\bs f) := \bs f^\top \bs\Phi\bs B\otop\bs B_\circ \, \bs f - \bs\lambda\!^\top \bs f \,\Bigr)
\end{equation}
with $\bs\lambda$ being the vector of Lagrange multipliers. 
Since the primal variable $\bs f$ is continuously differentiable, so the Lagrangian $\cal L$, and thus the minimum occurs where $\nabla_{\!\bs f}\,{\cal L} = \bs 0$, which leads to
\begin{equation*}\label{ProofLem1_4}
\bs\lambda = \left(\bs B\otop\bs B_\circ\bs\Phi + \bs\Phi\bs B\otop\bs B_\circ\right)\bs f^\opt.
\end{equation*}
Substituting $\bs f^\opt$ in \eqref{ProofLem1_3} and noting that both $\bs f^\opt$ and $\bs\lambda$ are entrywise non-negative, we obtain
\begin{equation}\label{ProofLem1_5}
\max_{\bs\lambda\succcurlyeq\bs 0}\;\, {\cal L}(\bs\lambda) =  \max_{\bs\lambda\succcurlyeq\bs 0}\; -\frac{1}{2}\,\bs \lambda\!^\top \bs f^\opt = 0\,.
\end{equation}
By the weak duality theorem, the minimum of the primal problem is greater than or equal to the maximum of the dual problem.
Thus, \eqref{ProofLem1_5} entails $\min_{\bs f \succcurlyeq \bs 0}\,\bs f^\top \bs\Phi\bs B\otop\bs B_\circ \bs f \!\geqslant 0$, which equally means $\bs x^{\!\top}\bs M_{\!\circ}\,\bs x \geqslant 0$. 

It remains to show that $\bs x^{\!\top}\bs M_{\!\circ}\,\bs x = 0$ only if $\bs x = 0$, which is equivalent to show that matrix $\bs M\otop\!+\bs M_{\!\circ}$ is positive definite.
Since $\bs x^{\!\top}(\bs M\otop\!+\bs M_{\!\circ})\,\bs x \!\geqslant 0$ already guarantees that $\bs M\otop\!+\bs M_{\!\circ}$ is positive semi-definite, it suffices to show that $\bs M\otop\!+\bs M_{\!\circ}$ has no zero eigenvalue.  
Let us assume it does, which implies the existence of an eigenvector $\bs\nu\neq\bs 0$ such that
\begin{equation}\label{ProofLem1_6}
(\bs M\otop\!+\bs M_{\!\circ})\bs\nu = \bs 0\,.
\end{equation}
Because $\bs M_{\!\circ}$ is the product of two positive definite matrices, $\bs\nu\neq\bs 0$ entails $\bs M_{\!\circ}\bs\nu\neq\bs 0$, which leads to 
$(\bs M_{\!\circ}\bs\nu)^{\!\top}\bs M_{\!\circ}\bs\nu + (\bs M\otop\bs\nu)^{\!\top}\bs M\otop\bs\nu > 0$.
Utilizing \eqref{ProofLem1_6} in the latter results in  
\begin{equation*}\label{ProofLem1_8}
\bs\nu^\top \left(\,\bs M\otop\! - \bs M_{\!\circ} \right)^2\,\bs\nu < 0
\end{equation*}
which is not true as $\smash{(\bs M\otop\! - \bs M_{\!\circ})^2}$ is a symmetric positive semi-definite matrix.
Therefore, $\bs M\otop\!+\bs M_{\!\circ}$ has no zero eigenvalue and so is symmetric positive definite.
\QEDA

\subsection*{Proof of Lemma~\ref{Lemma2}}

By definition of $\bs M_{\!\circ}$, we already have $\bs B_\circ\bs B\otop \bs M_{\!\circ}\,\bs x = \bs B_\circ\bs\Phi\bs B\otop\bs x$, which could easily be seen by substituting $\bs M_{\!\circ}$ from \eqref{Lyapunov1}.
Thus, to prove the claim, it suffices to show that for any vectors $\bs x$ and $\bs y$, equality $\bs B_\circ\bs y = \bs B_\circ\bs\Phi\bs B\otop\bs x$ entails $\bs y = \bs\Phi\bs B\otop\bs x$.
To this end, we utilize the properties of heat equations on undirected graphs (see Sec.~\ref{s:C3_heatUndirectedGraphs}).

Consider a thermal graph with reduced node-edge incidence matrix $\bs B_\circ$ and edge thermal diffusivity matrix $\bs\Phi$ and let the destination node be fixed at zero temperature. 
As the first scenario, let us envision $\bs y$ as the vector of heat fluxes through the branches, implying that $\bs B_\circ\bs y$ represents the vector of heat sources injected at the nodes (see \eqref{discrete4} and \eqref{discrete5} under constant heat sources.)
As the second scenario, envision $\bs x$ as the vector of temperatures at the nodes, implying that $\bs\Phi\bs B\otop\bs x$ represents the vector of heat fluxes through the branches and $\bs B_\circ\bs\Phi\bs B\otop\bs x$ represents the vector of heat sources injected at the nodes.

Assuming $\bs B_\circ\bs y = \bs B_\circ\bs\Phi\bs B\otop\bs x$ means that the thermal graph is charged by the same configuration of heat sources in both scenarios above. 
It follows that the vector of temperatures at the nodes are also the same as the Dirichlet Laplacian is positive definite in \eqref{discrete5}.
Hence, in both scenarios the vector of heat fluxes through the branches must be equal, because $\bs B_\circ$ has full row rank in \eqref{discrete4}.
This entails $\bs y = \bs\Phi\bs B\otop\bs x$, concluding the proof.
\QEDA

\subsection*{Proof of Lemma~\ref{Lemma3}}

Let us replace $\bs M_{\!\circ} + \bs M\otop$ by $2\bs M_{\!\circ} + (\bs M\otop \!- \bs M_{\!\circ})$.
Doing some matrix manipulation, we need to show that there exists such $1\leqslant\eta\leqslant3$ that for arbitrary vectors $\bs x$ and $\bs y$, 
\begin{equation}\label{ProofLem2_6}
\bs x^{\!\top} (\bs M\otop \!- \bs M_{\!\circ}) \,\bs y \leqslant
(\eta-2)\,\bs x^{\!\top} \bs M_{\!\circ}\,\bs y\,.
\end{equation}
To this end, it suffices to show 
$\left\vert\bs x^{\!\top} (\bs M\otop \!- \bs M_{\!\circ}) \,\bs y\right\vert \leqslant
\left\vert\bs x^{\!\top} \bs M_{\!\circ}\,\bs y\right\vert$,
which then makes the inequality \eqref{ProofLem2_6} true for $\eta=1$ in case of $\bs x^{\!\top} \bs M_{\!\circ}\,\bs y\leqslant0$, and for $\eta=3$ in case of $\bs x^{\!\top} \bs M_{\!\circ}\,\bs y>0$. 
This is equivalent to show that the following inequality holds:
\begin{equation*}
\bs x^{\!\top} (\bs M\otop \!- \bs M_{\!\circ})\,\bs y \bs y ^{\!\top}(\bs M_{\!\circ} \!- \bs M\otop)\,\bs x 
\leqslant \bs x^{\!\top} \bs M_{\!\circ} \,\bs y \bs y^{\!\top} \bs M\otop \bs x\,.
\end{equation*}
By little algebra, the latter can be rephrased as
\begin{equation*}
\bs x^{\!\top} (2\,\bs M_{\!\circ} \!- \bs M\otop)\,\bs y \bs y^{\!\top} \bs M_{\!\circ}\,\bs x \geqslant 0\,.
\end{equation*}
To prove the above inequality, it suffices to show that the minimum objective value in the following optimization problem is non-negative: 
\begin{equation*}\label{ProofLem2_1}
\begin{aligned}
\min\nolimits&_{\bs x , \bs y} & &\bs x^{\!\top} (2\,\bs M_{\!\circ} \!- \bs M\otop)\, \bs y \, \bs y^{\!\top} \bs M_{\!\circ} \bs x 
\\
\mathrm{s.t.}& & &\bs x^{\!\top} \bs M_{\!\circ} \bs x > 0 \hspace{5pt} , \hspace{5pt} \bs y^{\!\top} \bs M_{\!\circ} \bs y > 0
\end{aligned}
\end{equation*}
where the constraints are enforced in light of Lem.~\ref{Lemma1}.
The Lagrangian dual problem, with $\lambda_x$ and $\lambda_y$ as the Lagrange multipliers, is found as
\begin{equation*}\label{ProofLem2_2}
\begin{aligned}
\max_{\lambda_x,\lambda_y\geqslant0}\, 
\min_{\bs x,\bs y}\;\Bigl(\, {\cal L} := \bs x^{\!\top}  & 
(2\,\bs M_{\!\circ} \!- \bs M\otop)\, \bs y \, \bs y^{\!\top} \bs M_{\!\circ} \bs x
\\[-7pt] & 
- \lambda_x \, \bs x^{\!\top} \bs M_{\!\circ} \bs x - \lambda_y \, \bs y^{\!\top} \bs M_{\!\circ} \bs y         
\,\Bigr)
\end{aligned}
\end{equation*}
Imposing the first order conditions $\nabla_{\!\bs x}\,{\cal L} = \bs 0$ and $\nabla_{\!\bs y}\,{\cal L} = \bs 0$ leads to
\begin{equation*}\label{ProofLem2_3}
\begin{aligned}
\lambda_x \,(\bs M\otop\!+\bs M_{\!\circ})\,\bs x  &
= 2\,\bs M_{\!\circ} \bs y \, \bs y^{\!\top} \bs M_{\!\circ} \bs x + 2\,\bs M\otop \bs y \, \bs y^{\!\top} \bs M\otop \bs x  &
\\ & \hspace{30mm} 
- 2\,\bs M\otop \bs y \, \bs y^{\!\top} \bs M_{\!\circ} \bs x  &
\\
\lambda_y \,(\bs M\otop\!+\bs M_{\!\circ})\,\bs y &
= 2\,\bs M_{\!\circ} \bs x \, \bs x^{\!\top} \bs M_{\!\circ} \bs y + 2\,\bs M\otop \bs x \, \bs x^{\!\top} \bs M\otop \bs y  &
\\ & \hspace{30mm} 
- 2\,\bs M_{\!\circ} \bs x \, \bs x^{\!\top} \bs M\otop \bs y\,.  &
\end{aligned}
\end{equation*}
Let us plug these two equations into the Lagrangian ${\cal L}$ and utilize the identities 
$\bs x^{\!\top}\bs M\otop\bs y = \bs y^{\!\top}\bs M_{\!\circ}\bs x := a$ and 
$\bs x^{\!\top}\bs M_{\!\circ}\bs y = \bs y^{\!\top}\bs M\otop\bs x := b$ with $a$ and $b$ being scalars. 
One can easily confirm the following identities:  
\begin{gather*}
{\cal L} = (2\,a\,b - b^2) - \lambda_x \, \bs x^{\!\top} \bs M_{\!\circ} \bs x - \lambda_y \, \bs y^{\!\top} \bs M_{\!\circ} \bs y
\\
\lambda_x \,\bs x^{\!\top}(\bs M\otop\!+\bs M_{\!\circ})\,\bs x =
2\,(2\,a\,b - b^2)
\\
\lambda_y \,\bs y^{\!\top}(\bs M\otop\!+\bs M_{\!\circ})\,\bs y =
2\,(2\,a\,b - b^2).
\end{gather*}
Then by little algebra, the Lagrangian can be transformed to
\begin{equation*}\label{ProofLem2_4}
\begin{aligned}
{\cal L} = \:&
\frac{1}{4}\,\lambda_x\, \bs x{\!^\top\!} \bigl(\bs M\otop\!+\bs M_{\!\circ}\bigr)\bs x 
+ \frac{1}{4}\,\lambda_y\, \bs y{\!^\top\!} \bigl(\bs M\otop\!+\bs M_{\!\circ}\bigr) \bs y
\\ & \hspace{32.5mm}
- \lambda_x \, \bs x^{\!\top} \bs M_{\!\circ} \bs x - \lambda_y \, \bs y^{\!\top} \bs M_{\!\circ} \bs y
\\ = \:&
\frac{1}{4}\,\lambda_x\, \bs x{\!^\top\!} \bigl(\bs M\otop\!-3\,\bs M_{\!\circ}\bigr)\bs x 
+ \frac{1}{4}\,\lambda_y\, \bs y{\!^\top\!} \bigl(\bs M\otop\!-3\,\bs M_{\!\circ}\bigr) \bs y \,.
\end{aligned}
\end{equation*}
Since $\bs M_{\!\circ}\! - \!\bs M\otop$ is skew-symmetric, both $\bs x^{\!\top} (\bs M_{\!\circ} - \bs M\otop)\bs x$ and $\bs y^{\!\top} (\bs M_{\!\circ} - \bs M\otop)\bs y$ vanish.
In light of $\bs x^{\!\top} \bs M_{\!\circ} \bs x > 0$ and $\bs y^{\!\top} \bs M_{\!\circ} \bs y > 0$, the Lagrangian dual problem reads
\begin{equation*}\label{ProofLem2_5}
\max_{\lambda_x,\lambda_y\geqslant0}\; {\cal L} =  \max_{\lambda_x,\lambda_y\geqslant0}
-\frac{1}{2}\, \bigl( \lambda_x \, \bs x^{\!\top} \bs M_{\!\circ} \bs x + \lambda_y \, \bs y^{\!\top} \bs M_{\!\circ} \bs y \bigr) = 0\,.
\end{equation*}
This entails $\bs x^{\!\top} (2\,\bs M_{\!\circ} \!- \bs M\otop)\,\bs y \bs y^{\!\top} \bs M_{\!\circ}\,\bs x \geqslant 0$ by the weak duality theorem that the maximum of the dual problem provides a lower bound for the minimum of the primal problem. 
\QEDA

\subsection*{Proof of Theorem~\ref{maximum_throughput} (HD Throughput Optimality)(Cont.)}

To simplify the proof, we assume arrivals to be i.i.d. over timeslots.
For non-i.i.d. arrivals with stationary ergodic processes of finite mean and variance, a similar analysis can be done using $N$-slot Lyapunov drift \cite{Georgiadis06}, where the queue evolution \eqref{model5} is modified to
\begin{equation}\label{Nslot}
\bs q_\circ(n+N)=\bs q_\circ(n) + \!\!\sum_{k=n}^{n+N-1}\!\!\bs a_\circ(k) 
- \!\!\sum_{k=n}^{n+N-1}\!\!\bs B_\circ\bs f(k)\,. 
\end{equation}
One can view $N$ as the time required for the system to reach ``near steady state,'' noting that in the i.i.d. case, the steady state is reached on each and every timeslot, and so $N = 1$.

Back to the proof for i.i.d. arrivals, suppose that $\overline{\bs a_\circ\!}\,$ is interior to the network capacity region $\cal C$.
Thus, there exists an $\epsilon>0$ such that $\overline{\bs a_\circ\!}\, + \epsilon\bs 1 \in \cal C$.
Since the stationary randomized algorithm that generates $\bs{f^\prime}(n)$ is throughput optimal \cite{Georgiadis06}, it can stabilize the arrival $\overline{\bs a_\circ\!}\, + \epsilon\bs 1$ at each timeslot. 
The i.i.d. assumption on arrivals then leads to 
\begin{equation*}
\mathbb E\{\bs a_\circ \!-\! \bs B_\circ\bs{f^\prime}\} = 
\overline{\bs a_\circ\!}\, - (\overline{\bs a_\circ\!}\, + \epsilon\bs 1) =
- \epsilon\bs 1
\end{equation*}
implying that both $\bs a_\circ$ and $\bs f^{\bs\prime}$ reach their steady states on each and every timeslot.
Plugging this into the Lyapunov drift inequality~\eqref{th2_e} yields
\begin{equation}\label{proofTh2_1}
\mathbb E\bigl\{\Delta W\,\vert\bs q_\circ\bigr\} \leqslant 
- \eta\,\epsilon\bs 1^{\!\top}\mathbb E\{\bs M_{\!\circ}\} \,\bs q_\circ + \Gamma_{\!\max}\,.
\end{equation}
Let us assume that there exists a $\mu\!>0$, which is explored later, such that $\bs 1^{\!\top}\mathbb E\{\bs M_{\!\circ}\} \,\bs q_\circ \geqslant \mu\,\bs 1^{\!\top}\!\bs q_\circ$. 
Using this in the latter drift inequality leads to
\begin{equation*}\label{proofTh2_2}
\mathbb E\bigl\{\Delta W\,\vert\bs q_\circ\bigr\} \leqslant 
- \eta\,\mu\,\epsilon\,\bs 1^{\!\top}\!\bs q_\circ + \Gamma_{\!\max}\,.
\end{equation*}
Thus, $\mathbb E\{\Delta W\,\vert\bs q_\circ\}<0$ for any $\sum_i q_i > \Gamma_{\!\max}\bigl/(\eta\,\mu\,\epsilon)$.
Then in light of Theorem~2 in~\cite{Leonardi}, the queuing system is stable and so $\overline{\bs a_\circ\!}\,$ is in the HD stability region.
This implies that any arrival rate $\overline{\bs a_\circ\!}\,$ being interior to the network capacity region is stabilized by HD with any $\beta\in[0,1]$, meaning that HD is throughput optimal for all $\beta\in[0,1]$.

We now show that there exists such a $\mu\!>0$ that satisfies $\bs 1^{\!\top}\mathbb E\{\bs M_{\!\circ}\} \,\bs q_\circ \geqslant \mu\,\bs 1^{\!\top}\!\bs q_\circ$.
Let us temporarily ignore the expectation and solve the problem for $\bs M_{\!\circ}$.
The claim is trivial for $\bs q_\circ = \bs 0$, and so we assume $\bs q_\circ \neq \bs 0$.
Further, $\bs q_\circ$ represents the vector of queue occupancies on the wireless network that are always non-negative.
Let $\Vert\bs q_\circ \Vert_1$ represent the $\ell^1$ norm of $\bs q_\circ$, defined as the sum of all queue occupancies.
With no loss of generality, one may normalize $\Vert\bs q_\circ \Vert_1$ to one.
The problem can then be rephrased as finding a $\mu >0$ such that 
\begin{equation*}\label{proofTh2_3}
\min_{\Vert\bs q_\circ \Vert_1 = 1, \bs q_\circ \succcurlyeq \bs 0} \bs 1^{\!\top}(\bs M_\circ-\mu \bs I) \,\bs q_\circ \geqslant 0\,.
\end{equation*}
The latter is a standard linear programming problem. 
Using simplex method, the minimum lies on a vertex of the simplex, where the vertices of the simplex are the natural basis elements $\bs e_j : j=1,...,\vert{\cal V}\vert$. 
Thus, the $\mu$ is to be sought such that
\begin{equation*}\label{proofTh2_4}
\bs 1^{\!\top}\Bigl( (\bs M_\circ)_{:,j}-\mu \, \bs e_j  \Bigr) \geqslant 0\,.
\end{equation*}
This immediately leads to $\mu = \min_j \,\bs 1^{\!\top} \bs M_\circ \,\bs e_j$.

It remains to show that $\bs 1^{\!\top} \bs M_\circ \,\bs e_j > 0$ for every natural basis $\bs e_j$.
By Lem.~\ref{Lemma3}, there exists such a $1\leqslant\eta\leqslant3$ as
\begin{equation*}\label{proofTh2_5}
\eta\, \bs 1^{\!\top} \bs M_\circ \,\bs e_j \geqslant 
\bs 1^{\!\top} \bigl( \bs M\otop + \bs M_{\!\circ} \bigr) \,\bs e_j 
\end{equation*}
which implies 
$(\eta-1)\, \bs 1^{\!\top} \bs M_{\!\circ} \,\bs e_j \geqslant 
\bs 1^{\!\top} \bs M\otop \,\bs e_j$.
The right hand side is always positive by the next electrical circuit argument, which implies $(\eta-1)\, \bs 1^{\!\top} \bs M_\circ \,\bs e_j > 0$. 
The latter entails $\eta > 1$ and $\bs 1^{\!\top} \bs M_\circ \,\bs e_j > 0$ as we desired. 

To argue that $\bs 1^{\!\top} \bs M\otop \,\bs e_j = \bs e_j^\top \bs M_{\!\circ} \,\bs 1 > 0$, by substituting $\bs M_{\!\circ}$ from \eqref{Lyapunov1}, we need to show  
\begin{equation*}\label{proofTh2_6}
\bs e_j^{\!\top} (\bs B_\circ\bs B\otop){^{-1}}(\bs B_\circ\bs\Phi\bs B\otop)\,\bs 1 > 0\,.
\end{equation*}
Let us associate node-edge incidence matrix $\bs B_\circ$ with a resistive network and $\bs e_j$ with the vector of independent current sources attached to the nodes. 
Then the vector $\bs v := (\bs B_\circ\bs B\otop){^{-1}}\bs e_j$ reads the voltages induced at the nodes.
Since $\bs e_j$ implies that electrical current is injected into the network by a single current source at the node $j$, the resulting voltage at each node is non-negative ($\bs v \succcurlyeq \bs 0$). 
Further, the voltages at the node $j$ and at least at one of the nodes neighbor to ground (destination node) are always positive. 
On the other hand, the elements of each row of the Dirichlet Laplacian $\bs B_\circ\bs\Phi\bs B\otop$ sum to zero, except for those rows representing the nodes neighbor to ground, which always sum to a positive value. 
(Recall that $\bs B_\circ$ is obtained from $\bs B$ by discarding the row related to ground.)
This implies that in the vector $\bs u:=(\bs B_\circ\bs\Phi\bs B\otop)\,\bs 1$, the components related to the nodes neighbor to ground are positive, and others are zero. 
Considering the conditions of $\bs u$ and $\bs v$ together, we get $\bs e_j^\top \bs M_{\!\circ} \,\bs 1 = \bs v^\top \bs u > 0$.
Replacing $\bs M_{\!\circ}$ by $\mathbb E\{\bs M_{\!\circ}\}$, the same argument leads to $\mu = \min_j \,\bs 1^{\!\top} \mathbb E\{\bs M_{\!\circ}\}\, \bs e_j > 0$, which concludes the proof.
\QEDA

\subsection*{Proof of Lemma~\ref{Lemma4}}

Considering the maximum of $G(\bs f)$ subject to $\bs f\succcurlyeq \bs 0$, the Lagrangian dual problem is obtained as
\begin{equation*}\label{ProofLemma4_1a}
\min_{\bs\lambda\succcurlyeq\bs 0}\;\max_{\bs f}\,
\Bigl(\, {\cal L}(\bs\lambda,\bs f) := 
2\,\bs f^\top\bs B\otop\bs q_\circ - \bs f^\top \bs B\otop \bs B_\circ\bs f
+ \bs\lambda\!^\top \bs f \, \Bigr)
\end{equation*}
with $\bs\lambda$ being the vector of Lagrange multipliers. 
Using the first order condition $\nabla_{\!\bs f}\,{\cal L}=\bs 0$, we get 
\begin{equation}\label{ProofLemma4_2a}
2\,\bs B\otop \bs B_\circ\bs f^\opt = 2\,\bs B\otop\bs q_\circ + \bs\lambda \,.
\end{equation}
Plugging $\bs\lambda$ from \eqref{ProofLemma4_2a} into the Lagrangian ${\cal L}$ leads to
\begin{equation*}\label{ProofLemma4_3a}
\min \, {\cal L} = \bs f^{\opt\top} \bs B\otop \bs B_\circ\bs f^\opt .
\end{equation*}
By the weak duality theorem, the maximum of the primal problem is smaller than or equal to the minimum of the dual problem.
Further, the duality gap is zero as ${\cal L}(\bs\lambda,\bs f)$ is a convex functional, which leads to
\begin{equation*}\label{ProofLemma4_3ab}
\max \, G = \bs f^{\opt\top} \bs B\otop \bs B_\circ\bs f^\opt .
\end{equation*}
Substituting the latter into the $G$ functional entails
\begin{equation}\label{ProofLemma4_3ac}
\bs f^{\opt\top}\bs B\otop\bs q_\circ -
\bs f^{\opt\top} \bs B\otop \bs B_\circ\bs f^\opt = 0 \, .
\end{equation}
One can verify, on the other hand, that $\bs B\otop\!\bs B_\circ\bs f$ is an edge vector, in which
the entry corresponding to edge $ij$ reads
\begin{equation}\label{ProofLemma4_4a}
(\bs B\otop\bs B_\circ\bs f)_{ij} = (\bs B_\circ\bs f)_i - (\bs B_\circ\bs f)_j\,.
\end{equation}
Under the K\mbox{-}hop interference model, two wireless links that share a common node cannot be scheduled in the same timeslot. 
Thus, for each scheduled link $ij\in{\cal E}$, we get the net outflow for node $i$ as $(\bs B_\circ\bs f)_i = f_{ij}$.
When node $j$ is not the final destination, we get $(\bs B_\circ\bs f)_j = -f_{ij}$, and when it is, we get $(\bs B_\circ\bs f)_j = 0$. 
(Recall that $\bs B_\circ$ is a reduction of $\bs B$ by discarding the row related to the final destination.)
Using these identities in \eqref{ProofLemma4_4a}, we get
$(\bs B\otop\bs B_\circ\bs f)_{ij} = \vartheta_{ij} f_{ij}$ with $\vartheta_{ij}$ defined in \eqref{HD2}.
Substituting the latter into \eqref{ProofLemma4_3ac}, we obtain
\begin{equation*}\label{ProofLemma4_5a}
f_{ij}^\opt \bigl( q_{ij} - \vartheta_{ij} f_{ij} ^\opt \bigr) = 0 \,.
\end{equation*}
Considering the link constraints that $f_{ij}$ must be non-negative and at most equal to the link capacity yields
\[
f_{ij}^\opt\! = \min\{\, q_{ij}{^+} \!/ \vartheta_{ij} ,\: \mu_{ij} \}
\]
which follows $\smash{\widehat{f_{ij}}}$ in~\eqref{HD2} with $\beta=0$.
Next is to activate the links that contribute most to the $G$ maximization that directly leads to the max-weight scheduling~\eqref{HD4} alongside the HD weighting~\eqref{HD3} with $\beta=0$, concluding the proof.
\QEDA

\subsection*{Proof of Lemma~\ref{Lemma5}}

Consider $W(n) := \bs q_{\circ\!}(n)^{\!\top}\bs q_{\circ\!}(n)$ as the classical quadratic Lyapunov candidate and take expectation from the Lyapunov drift $\Delta W(n)=W(n+1)-W(n)$ to obtain 
\begin{equation}\label{ProofLem4-1}
\begin{aligned}
\mathbb E \{\Delta W\} = 
\mathbb E&\{\bs a_\circ\!-\!\bs B_\circ\bs f\}^{\!\top} \mathbb E\{\bs a_\circ\!-\!\bs B_\circ\bs f\!+2\,\bs q_\circ\}
\\&
-2\,\cov\{\bs B_\circ\bs f,\bs q_\circ\}
+\var\{\bs B_\circ\bs f\}
\\&
+2\,\cov\{\bs a_\circ,\bs q_\circ\!-\!\bs B_\circ\bs f\}
+\var\{\bs a_\circ\}
\end{aligned}
\end{equation}
where the equality holds at each timeslot and expectation is with respect to the randomness of arrivals, channel states and (possibly) routing decision.
Let $\bs g:=\bs a_\circ-\bs B_\circ\bs f+2\,\bs q_\circ$, sum over timeslots 0 until $\tau-1$, divide by $\tau$ and take a $\limsup$ of $\tau\to\infty$ from both sides of \eqref{ProofLem4-1} to obtain the following expected time average equation:
\begingroup
\setlength\abovedisplayskip{3pt} 
\begin{equation}\label{ProofLem4-2}
\begin{aligned}
\limsup_{\tau\to\infty}\;\frac{1}{\tau} \sum_{n=0}^{\tau-1}\;
\mathbb E&\{\bs a_\circ(n) - \bs B_\circ\bs f(n)\}^{\!\top} \mathbb E\{\bs g(n)\} =
\\[-6pt]
& + 2\,\,\overline{\!\cov\{\bs B_\circ\bs f,\bs q_\circ\}\!}\, 
-\overline{\var\{\bs B_\circ\bs f\}\!}\,
\\
& - 2\,\,\overline{\!\cov\{\bs a_\circ,\bs q_\circ\!-\!\bs B_\circ\bs f\}\!}\, 
-\overline{\var\{\bs a_\circ\}\!} 
\end{aligned}
\end{equation}
\endgroup
where we utilized $\limsup_{\tau\to\infty}(W(\tau)-W(0))/\tau=0$, as the routing policy stabilizes $\overline{\bs a_\circ\!}\,$ and so keeps $W(n)$ finite with probability 1 at each timeslot.

It remains to show that the left-hand side of \eqref{ProofLem4-2} vanishes.
Observe that $\bs g(n)$ is entrywise non-negative and finite.
Thus, there exist constant vectors $\bs g_{\min}$ and $\bs g_{\max}$ such that 
\begin{equation*}\label{ProofLem4-3}
\bs 0\preccurlyeq\bs g_{\min}\preccurlyeq\mathbb E\{\bs g(n)\}\preccurlyeq\bs g_{\max}.
\end{equation*}
Hence, the left-hand side of~\eqref{ProofLem4-2} is bounded from below to $(\,\overline{\bs a_\circ\!}\,-\bs B_\circ\overline{\bs f}\,)^{\!\top}\bs g_{\min}$ and from above to $(\,\overline{\bs a_\circ\!}\,-\bs B_\circ\overline{\bs f}\,)^{\!\top\!}\bs g_{\max}$.
Further, as $\overline{\bs a_\circ\!}\,$ is stabilized by the routing policy, the feasibility condition in~\eqref{hyperflow1} entails $\overline{\bs a_\circ\!}=\bs B_\circ\overline{\bs f}$, implying that the left-hand side of~\eqref{ProofLem4-2} vanishes.
\QEDA

\subsection*{Proof of Theorem~\ref{minimum_queue} (HD Minimum Delay)}

To simplify the proof, we assume arrivals are i.i.d. over timeslots, with the understanding that it can easily be modified to yield similar result for non-i.i.d. arrivals, using the $N$-slot analysis derived from \eqref{Nslot}.

Consider an arrival rate $\overline{\bs a_\circ\!}\,$ interior to the stability region of a $\cal D$-class routing policy, which we refer to it as ``generic''.
Let the timeslot quantities $\bs f(n)$ and $\bs q_\circ(n)$ be produced by such a generic routing policy.
If this generic routing policy also maximizes the $G$ functional \eqref{Dfunction3} at each slot $n$, by Assum.~\ref{ass2}, it will result in the same $\overline{Q}$ as that of HD policy at $\beta=0$.
Thus, we assume the $G$ obtained by the generic policy is not maximal.
Then for a sufficiently small $\epsilon>0$, there exists a routing algorithm (possibly unfeasible) that can stabilize the arrival $\overline{\bs a_\circ\!}\,+\epsilon\bs1$ while making $G(\bs f,\bs q_\circ,n)$ not less than that of the generic routing policy at each slot $n$. 
Let us refer to this algorithm as ``fictitious,'' as we do not intend to know how it really works.
To rest assure that such an algorithm exists, one may endow it with the ability of perfectly predicting all future events with no uncertainty.

Let $\bs{f^\prime}(n)$ represent the vector of link actual transmissions produced by the fictitious algorithm at slot $n$ given $\bs q_\circ(n)$.
Taking expectation from $G(\bs{f^\prime},\bs q_\circ,n) \geqslant G(\bs f,\bs q_\circ,n)$ and considering $\mathbb E\{\bs B_\circ{\bs f^\prime}\} = \mathbb E\{\bs B_\circ\bs f\}+\epsilon\bs 1$ due to the feasibility condition~\eqref{hyperflow1} and the i.i.d. arrivals, we obtain
\begin{align*}
2\,\epsilon\bs1\!^\top\mathbb E\{\bs q_\circ\} \geqslant \, & 
\:2\,\epsilon\,\bs1\!^\top\mathbb E\{\bs B_\circ\bs{f^\prime}\} - \epsilon^2\bs1\!^\top\bs1
\\
& + \bigl(\, 2\,\cov\{\bs B_\circ\bs f, \bs q_\circ\} - \var\{\bs B_\circ\bs f\} \,\bigr)
\\ 
& - \bigl(\, 2\,\cov\{\bs B_\circ\bs{f^\prime}, \bs q_\circ\} - \var\{\bs B_\circ\bs{f^\prime}\} \,\bigr)
\end{align*}
which holds for each timeslot.
Summing over timeslots 0 until $\tau-1$, dividing by $\tau$ and taking a $\limsup$ of $\tau\to\infty$ from both sides lead to the following expected time average inequality:
\begin{align*}
2\,\epsilon\,\bs 1\!^\top \,\overline{\!\bs q_\circ\!}\, \geqslant \, & 
\:2\,\epsilon\,\bs1\!^\top\hspace{5pt}\overline{\hspace{-6pt}(\bs B_\circ\bs{f^\prime})\hspace{-4pt}}\hspace{4pt} - \epsilon^2\bs1\!^\top\bs1
\\ 
& + \bigl(\, 2\,\,\overline{\!\cov\{\bs B_\circ\bs f,\bs q_\circ\}\!}\, - \overline{\var\{\bs B_\circ\bs f\}\!}\, \,\bigr)
\\  
& - \bigl(\, 2\,\,\overline{\!\cov\{\bs B_\circ\bs{f^\prime},\bs q_\circ\}\!}\, - \overline{\var\{\bs B_\circ\bs{f^\prime}\}\!}\, \,\bigr).
\end{align*}
Let us exploit Lem.~\ref{Lemma5} in the second and third lines and apply the identities $\cov\{\bs a_\circ + \epsilon\bs 1,\bs q_\circ\!-\!\bs B_\circ\bs{f^\prime}\} = \cov\{\bs a_\circ,\bs q_\circ\!-\!\bs B_\circ\bs{f^\prime}\}$ and $\var\{\bs a_\circ + \epsilon\bs1\} = \var\{\bs a_\circ\}$ to obtain
\begin{equation*}\label{ProofTh3_3}
\begin{aligned}
2\,\epsilon\,\bs 1\!^\top \,\overline{\!\bs q_\circ\!}\, \geqslant \, & 
\:2\,\epsilon\,\bs1\!^\top\hspace{5pt}\overline{\hspace{-6pt}(\bs B_\circ\bs{f^\prime})\hspace{-4pt}}\hspace{4pt} - \epsilon^2\bs1\!^\top\bs1
\\ &
+ 2\,\,\overline{\!\cov\{\bs a_\circ,\bs B_\circ\bs{f^\prime}\}\!}\, - 2\,\,\overline{\!\cov\{\bs a_\circ,\bs B_\circ\bs f\}\!}\;.
\end{aligned}
\end{equation*}
Since $\bs f$ produced by the generic routing policy is independent of arrival statistics, we get $\cov\{\bs a_\circ,\bs B_\circ\bs f\}=0$.
Replacing $\bs 1\!^\top \,\overline{\!\bs q_\circ\!}\,$ by the $\overline Q$ expression as defined in \eqref{average3}, we then obtain
\begin{equation}\label{ProofTh3_4}
2\,\epsilon\,\overline Q \, \geqslant \, 2\,\epsilon\,\bs1\!^\top\hspace{5pt}\overline{\hspace{-6pt}(\bs B_\circ\bs{f^\prime})\hspace{-4pt}}\hspace{4pt}
+ 2\,\,\overline{\!\cov\{\bs a_\circ,\bs B_\circ\bs{f^\prime}\}\!}\,  - \epsilon^2\bs1\!^\top\bs1\,.
\end{equation}

Consider this time HD policy at $\beta=0$ with the timeslot quantities of $\bs q_\circ^\star(n)$ and $\bs f^\star(n)$.
Let again $\bs{f^\prime}(n)$ be produced by the fictitious algorithm at each slot $n$ to stabilize the arrival $\overline{\bs a_\circ\!}\,+\epsilon\bs1$, but this time, given $\bs q_\circ^\star(n)$.
In light of Lem.~\ref{Lemma4}, $G(\bs{f^\prime}\!,\bs q_\circ^\star,n)\leqslant G(\bs f^\star\!,\bs q_\circ^\star,n)$ at each slot $n$. 
Performing the similar steps of taking expectation, exploiting $\mathbb E\{\bs B_\circ\bs{f^\prime}\} = \mathbb E\{\bs B_\circ\bs f^\star\}+\epsilon$, translating the results into the expected time average form, using the fact that $\cov\{\bs a_\circ, \bs B_\circ\bs f^\star\} = 0$ as $\bs f^\star$ is independent of arrival statistics, and applying Lem.~\ref{Lemma5} by knowing that HD policy is throughput optimal and so stabilizes $\overline{\bs a_\circ\!}\,$, we obtain
\begin{equation}\label{ProofTh3_5}
2\,\epsilon\,\overline{Q^\star\!} \, \leqslant \, 2\,\epsilon\,\bs1\!^\top\hspace{5pt}\overline{\hspace{-6pt}(\bs B_\circ\bs{f^\prime})\hspace{-4pt}}\hspace{4pt} 
+ 2\,\,\overline{\!\cov\{\bs a_\circ,\bs B_\circ\bs{f^\prime}\}\!}\, - \epsilon^2\bs1\!^\top\bs1\,.
\end{equation}
Comparing \eqref{ProofTh3_4} and \eqref{ProofTh3_5} along with $\epsilon>0$ lead to $\overline{Q^\star\!} \leqslant \overline{Q}$.
This means the average network delay under HD policy with $\beta=0$ remains less than or equal to that under any other $\cal D$-class routing policy, which was called ``generic'' here.
\QEDA

\subsection*{Proof of Theorem~\ref{prop_fluid_model1} (HD Fluid Model)}

The proof follows the exact same line of argument proposed in \cite[Theorem~2.3.2]{Dai99} and \cite[Proposition~4.12]{Bramson08}.
\QEDA

\subsection*{Proof of Theorem~\ref{prop_fluid_model2} (Wireless Network Thermodynamics)}

Let $\bs q_\circ^\star(t)$ and $\bs f^\star(t)$ denote the HD fluid model variables.
Consider the continuous-time Lyapunov function 
\begin{equation*}\label{ProofTh5_1}
Y(t) := \bigl(\bs q_\circ^\star(t) - \bs q_\circ^\rf \,\bigr)^{\!\top} \;\overline{\!\bs M_{\!\circ}\!}\; \bigl(\bs q_\circ^\star(t) - \bs q_\circ^\rf \,\bigr)
\end{equation*}
where $\overline{\bs M}_{\!\circ} = \bigl(\bs B_\circ\bs B\otop\bigr){^{-1}}\bs B_\circ\,\overline{\bs\Phi}\,\bs B\otop$ represents the time average expectation of matrix $\bs M_{\!\circ}(n)$ as defined in \eqref{Lyapunov1}.
Taking time derivative from $Y(t)$, we obtain
\begin{equation*}\label{ProofTh5_2}
\dot Y(t) = \bs{\dot q}_\circ^\star(t)^{\!\top} 
\bigl(\;\overline{\!\bs M_{\!\circ}\!}^\top + \;\overline{\!\bs M_{\!\circ}\!}\;\bigr)
\bigl(\bs q_\circ^\star(t) - \bs q_\circ^\rf\,\bigr).
\end{equation*}
Exploiting Lem.~\ref{Lemma3} in the latter leads to
\begin{equation}\label{ProofTh5_3}
\dot Y(t) \leqslant \eta\;\bs{\dot q}_\circ^\star(t)^{\!\top} 
\;\overline{\!\bs M_{\!\circ}\!}\; \bigl(\bs q_\circ^\star(t) - \bs q_\circ^\rf\,\bigr)
\end{equation}
for an $1\leqslant\eta\leqslant3$.
As a positive coefficient, $\eta$ has no impact on the Lyapunov argument and can simply be omitted, but for the sake of consistency we prefer to keep it in here.

To find an appropriate expression for $\bs{\dot q}_\circ^\star(t)$, let us begin by plugging \eqref{fluid4a} in \eqref{fluid1} and taking time derivative to obtain
\begin{equation}\label{ProofTh5_4}
\bs{\dot q}_\circ^\star(t)= \overline{\bs a_\circ\!}\, - \bs B_\circ\bs{\dot f}{^{\star\tot}}(t)\,.
\end{equation}
Note in~\eqref{fluid4} that the entry of $\smash{\widehat{\bs f(}}t)$ corresponding to link $ij$ specifies the number of packets the link will send {\it per unit time} if it is activated at time~$t$. 
Then $\bs f(t)$ identifies the vector of rate of actual transmissions realized at time~$t$.
Assume now that the entry of $\bs f(t)$ corresponding to link $ij$ at time $t$ is equal to $x\!\geqslant0$, i.e., at time $t$ the link transmits $x$ number of packets per unit time.
Then it should be obvious that the same entry of $\bs{\dot f}{^\tot}(t)$ at time $t$ must also be equal to $x$.
In light of $\lim_{\delta\to0}\bs f^\tot(t+\delta) = \bs f^\tot(t) + \delta\bs f(t)$, this can be explained more formally by the classical definition of limit as
\begin{equation*}\label{ProofTh5_5}
\bs{\dot f}{^\tot}(t) = \lim_{\delta\to0}\,\frac{\bs f^\tot(t+\delta)-\bs f^\tot(t)}{\delta} = \bs f(t)\,.
\end{equation*}
Further, \eqref{discrete8}--\eqref{discrete9} imply $\overline{\bs a_\circ\!}=\LL{_\circ^\rf} \bs q_\circ^\rf = \bs B_\circ\bs f^\rf\!$.
Exploiting these latter identities in \eqref{ProofTh5_4} yields 
\begin{equation}\label{ProofTh5_6}
\bs{\dot q}_\circ^\star(t) = \bs B_\circ\bs f^\rf\! - \bs B_\circ\bs f^\star(t)\,.
\end{equation}

Returning to the Lyapunov argument, let us substitute \eqref{ProofTh5_6} in \eqref{ProofTh5_3} and utilize equality~\eqref{Lem1} in Lem.~\ref{Lemma2} to obtain 
\begin{equation*}\label{ProofTh5_7}
\eta^{-1}\,\dot Y(t) \leqslant \bigl( \bs f^\rf - \bs f^\star(t) \bigr)^{\!\top}
\,\overline{\bs\Phi}\,\bs B\otop
\bigl(\bs q_\circ^\star(t) - \bs q_\circ^\rf\bigr)\,.
\end{equation*}
Multiplying both sides by two, adding and subtracting the term $\bs f^\star(t)^{\!\top}\!\bs f^\star(t) + \bs f^{\rf\top}\!\bs f^\rf$ on the left-hand side, and recasting the terms lead to
\begin{subequations}\label{ProofTh5_8}
\begin{align}
\!\!\!2\,\eta^{-1}\,\dot Y(t) \leqslant &
- \bigl(2\,\bs f^\star(t)^{\!\top}\,\overline{\bs\Phi}\,\bs B\otop\bs q_\circ^\star(t) - \bs f^\star(t)^{\!\top}\!\bs f^\star(t)\bigr)
\label{ProofTh5_8a}
\\ &
+\bigl(2\,\bs f^{\rf\top}\,\overline{\bs\Phi}\,\bs B\otop\bs q_\circ^\star(t) - \bs f^{\rf\top}\!\bs f^\rf\bigr)
\label{ProofTh5_8b}
\\ & 
-\bigl(2\,\bs f^{\rf\top}\,\overline{\bs\Phi}\,\bs B\otop\bs q^\rf_\circ - \bs f^{\rf\top}\!\bs f^\rf\bigr)
\label{ProofTh5_8c}
\\ & 
+\bigl(2\,\bs f^\star(t)^{\!\top}\,\overline{\bs\Phi}\,\bs B\otop\bs q^\rf_\circ - \bs f^\star(t)^{\!\top}\!\bs f^\star(t)\bigr).
\label{ProofTh5_8d}
\end{align}
\end{subequations}
Characterizing \eqref{ProofTh5_8a} and \eqref{ProofTh5_8b} on the wireless network given $\bs q_\circ^\star(t)$, they respectively read $-D(\bs f^\star\!,\bs q_\circ^\star,t)$ and $D(\bs f^\rf\!,\bs q_\circ^\star,t)$.
Under Assum.~\ref{ass2} and in light of the HD fluid equations \eqref{fluid4} and \eqref{fluid51}, the immediate result of Th.~\ref{prop_fluid_model1} is that at each time $t$, HD fluid limit maximizes the $D$ functional compared to any alternative forwarding that satisfies wireless network constraints.
The $\bs f^\rf$ obviously meets the directionality constraints due to the structure of the reference thermal model. 
It also meets the capacity constraints due to Assum.~\ref{ass1}.
Hence, $\eqref{ProofTh5_8a} + \eqref{ProofTh5_8b} \leqslant0$ which leads to 
\begin{subequations}\label{ProofTh5_9}
\begin{align}
2\,\eta^{-1}\,\dot Y(t) \leqslant &
-\bigl(2\,\bs f^{\rf\top}\,\overline{\bs\Phi}\,\bs B\otop\bs q^\rf_\circ - \bs f^{\rf\top}\!\bs f^\rf\bigr)
\label{ProofTh5_9a}
\\ & 
+\bigl(2\,\bs f^\star(t)^{\!\top}\,\overline{\bs\Phi}\,\bs B\otop\bs q^\rf_\circ - \bs f^\star(t)^{\!\top}\!\bs f^\star(t)\bigr).
\label{ProofTh5_9b}
\end{align}
\end{subequations}
We now characterize \eqref{ProofTh5_9a} and \eqref{ProofTh5_9b} on the reference thermal model and let
\begin{equation*}\label{ProofTh5_10}
H(\bs f) := 2\,\bs f^\top\,\overline{\bs\Phi}\,\bs B\otop\bs q^\rf_\circ - \bs f^\top\!\bs f\,.
\end{equation*}
It can be shown that given $\bs q^\rf_\circ$, the maximum of $H$ occurs at $\bs f=\bs f^\rf$ produced by heat flow.
To see this, rephrase $H$ as
\begin{equation*}\label{ProofTh5_11}
H(\bs f) = \sum\nolimits_{ij\in\cal E}
2\,\,\overline{\!\phi_{ij}\!}\;q_{ij}^\rf f_{ij} - \bigl(f_{ij}\bigr){^2}
\end{equation*}
where directionality constraints entail $f_{ij}\!\geqslant0$.
To maximize $H$, one then needs to assign $f_{ij}=0$ if $q_{ij}^\rf\!\leqslant0$, and 
$f_{ij}\!=\,\overline{\!\phi_{ij}\!}\;q_{ij}^\rf$ otherwise.
Putting this back in a matrix form, we arrive at the same expression as $\bs f^\rf$ in \eqref{discrete8}.
Further, the maximizing $\bs f$ is unique by the reason that a given $\bs q^\rf_\circ$ leads to a unique $\bs B\otop\bs q^\rf_\circ$, and so to unique $q_{ij}^\rf$ components, as the matrix $\bs B_\circ$ has full row rank.
From $H(\bs f^\rf) \geqslant H(\bs f^\star(t))$, we then obtain $\eqref{ProofTh5_9a} + \eqref{ProofTh5_9b} \leqslant0$, which by $1\leqslant\eta\leqslant3$ yields $\dot Y(t) \leqslant0$.

Let $\Omega$ be the largest invariant set in the set of all $\bs q_\circ^\star(t)$ trajectories for which $\dot Y(t)=0$. 
Since $Y(t)$ is a non-negative and radially unbounded function with $\dot Y(t) \leqslant0$, LaSalle's invariance principle states that every trajectory $\bs q_\circ^\star(t)$ asymptotically converges to $\Omega$. 
It remains to show that $\Omega$ contains only the trivial trajectory of $\bs q_\circ^\star = \bs q_\circ^\rf$. 
If $\dot{Y}=0$, then \eqref{ProofTh5_9} entails $H(\bs f^\star(t)) = H(\bs f^\rf)$.
We previously showed as well that $\bs f^\rf$ maximizes $H$ and is unique, which implies 
\begin{equation}\label{ProofTh5_12}
\bs f^\star = \bs f^\rf.
\end{equation}
The intentionally dropped time variable $(t)$ in \eqref{ProofTh5_12} emphasizes that $\bs f^\star(t)$ turns to be stationary by converging to $\bs f^\rf$, which in turn entails $\bs q^\star(t)$ being converged to a stationary $\bs q^\star$ too. 

Given $\bs q^\star$, the equality \eqref{ProofTh5_12} entails that $\bs f^\rf$ must maximize $D$, which implies $f^\rf_{ij}=0$ if $q_{ij}^\star\!\leqslant0$, and $f^\rf_{ij}\!=\,\overline{\!\phi_{ij}\!}\;q_{ij}^\rf$ otherwise. 
In a matrix form, this is equivalent to $\bs f^\rf = \overline{\bs\Phi} \, \max\bigl\{ \bs 0 , \: \bs B\otop \bs q_\circ^\star \bigr\}$.
Putting the latter against \eqref{discrete8} leads to
\begin{equation}\label{ProofTh5_13}
\max\bigl\{ \bs 0 , \: \bs B\otop \bs q_\circ^\star \bigr\} = \max\bigl\{ \bs 0 , \: \bs B\otop \bs q_\circ^\rf \bigr\}.
\end{equation}
Consider a directed edge $ad$ with its head at the destination node, which has zero queue on the wireless network and zero temperature on the reference thermal model. 
By \eqref{ProofTh5_13}, $q_a^\star$ and $\bs q_a^\rf$ must be equal.
Repeating this argument eventually yields $(\bs q_\circ^\star)^+ \!= (\bs q_\circ^\rf)^+$, as any node with positive queue (resp. positive temperature) on the wireless network (resp. on the reference thermal model) must be connected to the destination node $d$ through a directed path. 
Further, observe that $\bs q_\circ^\star\!\succcurlyeq0$, as queues cannot be negative in a wireless network, and $\bs q_\circ^\rf\!\succcurlyeq0$, as temperatures cannot fall below zero in a thermal system with no negative heat source.
Thus, $\bs q_\circ^\star = \bs q_\circ^\rf$, which together with \eqref{ProofTh5_12} conclude the proof.
\QEDA

\subsection*{Proof of Theorem~\ref{prop_fluid_model3} (Nonlinear Dirichlet Principle)}

One can verify, by the $\LL_\circ$ structure in \eqref{discrete7}, that
\begin{equation*}\label{ProofTh6_1}
\vec E_D(\bs q_\circ) = \frac{1}{2}\, 
\bigl( \bs q\otop\bs B_\circ \bigr){^+}
\,\diag(\bs\sigma)
\bigl( \bs B\otop\bs q_\circ\bigr){^+}\!
- \bs q\otop\bs a_\circ\,
\end{equation*}
where each entry of $\bs B\otop\bs q_\circ$ represents temperature-difference along the corresponding edge.
Let $\bs q_\circ^*$ be the $\vec E_D(\bs q_\circ)$ minimizing solution and let us rearrange and partition $\bs B\otop\bs q_\circ^*$ into positive, zero and negative components.
Accordingly, $\bs B_\circ$ gets partitioned into $\bs B\opos$, $\bs B\onul$ and $\bs B\oneg$, which respectively contain the incidence information of edges with positive, zero and negative values in $\bs B\otop\bs q_\circ^*$.
Likewise, $\bs\sigma$ gets partitioned into $\bs\sigma\opos$, $\bs\sigma\onul$ and $\bs\sigma\oneg$.
Then at $\bs q_\circ=\bs q_\circ^*$, we obtain
\begin{subequations}
\begin{align}
\vec E_D(\bs q_\circ^*) = & -\bs q_\circ^{*\top}\! \bs a_\circ
+\frac{1}{2}\,\bigl(\bs q_\circ^{*\top}\! \bs B\opos \bigr){^+} \,\diag(\bs\sigma\opos) 
\bigl( \bs B\opos^\top \bs q_\circ^* \bigr){^+}
\nonumber
\\[-3pt]&
+\frac{1}{2}\,\bigl(\bs q_\circ^{*\top}\! \bs B\onul \bigr){^+} \,\diag(\bs\sigma\onul) 
\bigl( \bs B\onul^\top \bs q_\circ^* \bigr){^+}
\label{ProofTh6_2b}
\\[-2pt]&
\hspace{5mm}+\frac{1}{2}\,\bigl(\bs q_\circ^{*\top}\! \bs B\oneg \bigr){^+} \,\diag(\bs\sigma\oneg) 
\bigl( \bs B\oneg^\top \bs q_\circ^* \bigr){^+}.
\label{ProofTh6_2c}
\end{align}
\end{subequations}
Observe that \eqref{ProofTh6_2b} is strongly zero due to the $(\cdot){^+}$ operation.
On the other hand, \eqref{ProofTh6_2c} vanishes since $\bs B\onul^\top \bs q_\circ^*=\bs 0$.
In light of $(\bs B\opos^\top \bs q_\circ^*){^+} = \bs B\opos^\top \bs q_\circ^*$, we then obtain
\begin{equation}\label{ProofTh6_3}
\vec E_D(\bs q_\circ^*) = 
\frac{1}{2}\, \bs q_\circ^{*\top}\! \bs B\opos \,\diag(\bs\sigma\opos) \bs B\opos^\top \bs q_\circ^*
- \bs q_\circ^{*\top}\! \bs a_\circ\,.
\end{equation}
Since $\bs a_\circ$ is feasible, each nonzero heat source connects to the sink through at least one directed path. 
Thus, under any flow that keeps $\bs q_\circ$ entrywise finite, the edges with positive temperature-difference build a connected graph with the node~$d$.
On the other hand, $\bs q_\circ^*$ is entrywise finite as it minimizes $\vec E_D(\bs q_\circ)$, and so the corresponding edges in $\bs B\opos$ build a connected graph with the node~$d$.
This implies that $\bs B\opos \,\diag(\bs\sigma\opos) \bs B\opos^\top$ is a positive definite matrix.
Thus, the functional $\frac{1}{2}\, \bs q\otop \bs B\opos \,\diag(\bs\sigma\opos) \bs B\opos^\top \bs q_\circ - \bs q\otop \bs a_\circ$ is strictly convex in $\bs q_\circ$ 
and so finds its minimum at the critical point, where its first order variation with respect to $\bs q_\circ$ vanishes.
Comparing this with \eqref{ProofTh6_3}, it turns out that the minimizing $\bs q_\circ^*$ must satisfy 
\begin{equation}\label{ProofTh6_4}
\bs a_\circ = \bs B\opos \,\diag(\bs\sigma\opos) \,\bs B\opos^\top \bs q_\circ^*\,.
\end{equation}
Utilizing $\bs B\opos^\top \bs q_\circ^* \!= (\bs B\opos^\top \bs q_\circ^*){^+}$ and $(\bs B\oneg^\top \bs q_\circ^*){^+} \!= (\bs B\onul^\top \bs q_\circ^*){^+} \!= \bs 0$, one can rephrase \eqref{ProofTh6_4} as
\begin{equation*}\label{ProofTh6_5}
\bs a_\circ = \bs B_\circ \,\diag(\bs\sigma) \bigl(\bs B\otop\bs q_\circ^* \bigr){^+} = \LL_\circ \bs q_\circ^*
\end{equation*}
that recovers the nonlinear Poisson equation~\eqref{Poisson3} at $\bs q_\circ^*$.
Further, $\bs q_\circ^*$ is unique as it minimizes the strictly convex functional
$\frac{1}{2}\, \bs q\otop \bs B\opos \,\diag(\bs\sigma\opos) \bs B\opos^\top \bs q_\circ - \bs q\otop \bs a_\circ$, which concludes the proof.
\QEDA

\subsection*{Proof of Theorem~\ref{prop_fluid_model5} (Nonlinear Thomson Principle)}

Consider \eqref{Poisson7} as the primal optimization problem and let us construct its Lagrangian dual problem as
\begin{equation*}\label{ProofTh7_1}
\max_{\bs\lambda}\;\min_{\bs f\succcurlyeq\bs 0}\,
\Bigl(\, {\cal L}(\bs\lambda,\bs f) := \bs f^\top \,\diag(\bs\sigma)^{-1} \bs f + 2\,\bs\lambda\!^\top
\bigl(\bs a_\circ - \bs B_\circ\bs f \bigr)   \Bigr)
\end{equation*}
where $\bs\lambda\succcurlyeq0$ is the vector of Lagrange multipliers. 
From the first order condition $\nabla_{\!\bs f}\,{\cal L}=\bs 0$, we get $\bs f^\opt\!=\diag(\bs\sigma)\,\bs B\otop\bs\lambda$.
Then enforcing the constraint $\bs f^\opt\!\succcurlyeq\bs 0$ leads to $\bs B\otop\bs\lambda\succcurlyeq\bs 0$, which is equivalent to
$\bs B\otop\bs\lambda = (\bs B\otop\bs\lambda){^+}$.
Thus, we obtain
\begin{equation}\label{ProofTh7_2}
\bs f^\opt \!= \diag(\bs\sigma)\,(\bs B\otop\bs\lambda){^+}.
\end{equation}
Plugging this $\bs f^\opt$ into the Lagrangian ${\cal L}$ and utilizing the structure of $\LL_\circ$ in \eqref{discrete7}, we obtain
\begin{equation*}\label{ProofTh7_3}
{\cal L}(\bs\lambda) = -\bs\lambda\!^\top \LL_\circ \,\bs\lambda + 2\,\bs\lambda\!^\top \bs a_\circ\,.
\end{equation*}
Then the dual problem reads $\max_{\bs\lambda}\,{\cal L}(\bs\lambda)$, which is equivalent to the following minimization problem:
\begin{equation}\label{ProofTh7_4}
\min_{\bs\lambda}\; \frac{1}{2}\,\bs\lambda\!^\top \LL_\circ \,\bs\lambda - \bs\lambda\!^\top \bs a_\circ\,. 
\end{equation}
Further, as $\bs f\succcurlyeq\bs 0$ makes a convex set and ${\cal L}(\bs\lambda,\bs f)$ is a convex function, the duality gap is zero, and so both the primal and dual problems result in the same optimal solution.

Comparing \eqref{ProofTh7_4} with the nonlinear Dirichlet equation \eqref{Poisson4}, it remains to show that the Lagrangian multipliers $\bs\lambda$ are identical to the node temperatures $\bs q_\circ$.
In \eqref{ProofTh7_2}, multiplying both sides by $\bs B_\circ$ and using the $\LL_\circ$ expression, we obtain 
\begin{equation}\label{ProofTh7_5}
\LL_\circ\bs\lambda = \bs B_\circ\bs f^\opt = \bs a_\circ
\end{equation}
where the second equality comes from the constraint in the primal problem \eqref{Poisson7}.
Further, by Th.~\ref{prop_fluid_model3}, the nonlinear Poisson equation $\LL_\circ \bs q_\circ = \bs a_\circ$ has a unique solution.
Putting this against \eqref{ProofTh7_5} leads to $\bs\lambda=\bs q_\circ$, which concludes the proof. 
\QEDA

\subsection*{Proof of Theorem~\ref{prop_fluid_model4} (HD Minimum Routing Cost)}

It was shown by Th.~\ref{prop_fluid_model3} that if $\overline{\bs a_\circ\!}\,$ is feasible, then under the nonlinear heat equations~\eqref{discrete8}--\eqref{discrete9}, the stationary value of the nonlinear Dirichlet energy~$\vec E_D(\bs q_\circ)$ is strictly minimized. 
It was shown by Th.~\ref{prop_fluid_model5}, on the other hand, that minimizing $\vec E_D(\bs q_\circ)$ is equivalent to minimizing the stationary value of total energy dissipation~$\vec E_R(\bs f)$ on the graph. 
Then the proof immediately follows from Th.~\ref{prop_fluid_model2} which states that under a stabilizable arrival rate $\overline{\bs a_\circ\!}\,$, HD fluid model complies with the nonlinear heat equations~\eqref{discrete8}--\eqref{discrete9}.
Note that if $\overline{\bs a_\circ\!}\,$ is stabilizable, i.e., it satisfies condition \eqref{hyperflow1}, then its feasibility is trivial in the sense of Def.~\ref{def_feasible}.
\QEDA

\subsection*{Proof of Theorem~\ref{Pareto_performance} (HD Pareto Optimality)}

Observe that HD policy minimizes $\overline Q$ at $\beta\!=\!0$, minimizes $\overline R$ at $\beta\!=\!1$, and changes weight on these two objectives by altering $\beta$ between 0 and 1.
In fact, HD transforms the two objectives of minimizing $\overline Q$ and $\overline R$ into an aggregated objective function by multiplying each objective function by a weighting factor and summing up the two weighted objective functions.
Further, the weighted sum is a convex combination of objectives as the sum of weighting factors $\beta$ and $1\!-\!\beta$ equals 1.  
Under the assumption that the region $(\overline Q\,,\overline R)$ has a convex Pareto boundary, the proof then follows by the fact that the entire boundary can be reached using the weighted-sum method~\cite{Stadler84, Koski88, Kim05} that changes  the weight on the weighted convex combination of the two objective functions.
\QEDA


\bibliographystyle{plain}
\bibliography{RezaBIB3}


\section*{Appendix B}
\label{s:appendixB}

\subsection*{Graph Laplacian}

Consider a {\it connected,} weighted graph with set of nodes~$\cal V$, set of undirected edges~$\cal E$, node-edge incidence matrix $\bs B$, edge weight vector $\bs\sigma$ and the Laplacian $\bs L := \bs B\,\diag(\bs\sigma)\,\bs B^{\!\top}$.
One can verify that $\bs x^{\!\top}\bs L\bs x = \sum_{ij\in\cal E}\sigma_{ij}(x_i - x_j)^2 \geqslant 0$, which means $\bs L$ is {\it positive semi-definite.}
Observe that $\bs L\bs 1 = \bs 0$, which implies that $\bs 1$ is an eigenvector corresponding to the smallest eigenvalue $\lambda_1 = 0$.
For the second smallest eigenvalue $\lambda_2$, let $\bs\nu$ be the eigenvector orthogonal to $\bs 1$.
Thus, ${\bs \nu^{\!\top}\bs 1=0}$ and $\lambda_2 = \bs \nu^{\!\top}\bs L\bs \nu = \sum_{ij\in\cal E}\sigma_{ij}(\nu_i - \nu_j)^2$.
Assume $\lambda_2 = 0$.
Since the graph is connected, there exists a path between every two nodes, which enforces $\bs \nu = c\bs 1$ for a constant $c$.
This contradicts $\bs \nu^{\!\top}\bs 1=0$, and so $\lambda_2$ must be positive. 

Let $\bs L^\dag$ be the Moore-Penrose pseudoinverse of $\bs L$.
Both $\bs L$ and $\bs L^\dag$ have the same eigenvectors, while two corresponding eigenvalues are reciprocals of each other, except that 1 replaces the zero eigenvalue of $\bs L$. 
One can verify that 
\begin{gather*}
\bs L^\dag = \Big( \bs L + \frac{1}{n}\,\bs1\!^\top\bs 1 \Big)^{-1} \!- \frac{1}{n}\,\bs1\!^\top\bs 1 
\;\text{ with }\;
n:=\vert{\cal V}\vert.
\end{gather*}
Also $\bs L^\dag$ enjoys the structural property of $\bs L^\dag\bs 1 = \bs 0$.  

The Dirichlet Laplacian $\bs L_\circ := \bs B_\circ\,\diag(\bs\sigma)\,\bs B\otop$ is made from $\bs L$ by discarding the entries corresponding to a {\it reference} node $d$.
Let ${\cal E} = {\cal E}_1 + {\cal E}_\circ$, where ${\cal E}_\circ$ is the set of edges with one end connected to node $d$.
One can verify that $\forall\,\bs x \neq \bs 0$, 
\begin{equation*}
\bs x^{\!\top}\bs L_\circ\bs x = \textstyle\sum_{ij\in{\cal E}_1}\!\sigma_{ij}(x_i - x_j)^2 + \textstyle\sum_{id\in{\cal E}_\circ}\!\sigma_{id}\,x_i^2 > 0
\end{equation*}
which means $\bs L_\circ$ is {\it positive definite} and so {\it invertible.}
Further, $\bs L_\circ\bs 1 = \bs y$ is a nonzero vector with non-negative coordinates, where $y_i = 0$ if $id\in{\cal E}_1$ and $y_i > 0$ if $id\in{\cal E}_\circ$.

Sometimes, it is misunderstood that $\bs L_\circ$ carries the same eigenvalues as $\bs L$ but the zero, which is not true.

\end{document}